\documentclass[a4paper,11pt]{article}

\usepackage{jcappub} 

\usepackage[T1]{fontenc} 
\usepackage{amssymb,amsmath,amsfonts,amsbsy,graphicx}
\usepackage{color}
\usepackage{psfrag}
\usepackage{stackrel}
\usepackage{bm}
\usepackage{tikz}
\usetikzlibrary{positioning}
\usepackage{siunitx}
\usetikzlibrary{intersections}
\usepackage{url}
\usepackage{svg}
\usepackage{comment}
\usepackage{here}

\newcommand\lcdm{$\Lambda$CDM }

\newcommand{\Msun}{M_\odot}

\newcommand{\Msunhinv}{M_\odot /h}
\newcommand{\vir}{\mathrm{vir}}

\newcommand{\mr}{\mathrm}
\newcommand{\mrd}{\mathrm{d}}

\newcommand{\romaii}{i\hspace{-1.2pt}i}
\newcommand{\romaiii}{i\hspace{-1.2pt}i\hspace{-1.2pt}i}
\newcommand{\romaiv}{i\hspace{-1.2pt}v}

\title{\boldmath The Contribution of Low-Mass Minihalos to the 21cm Forest}

\author[a]{Genki Naruse,\note{Corresponding author.}}
\author[b]{Yuxiang Qin,}
\author[b]{Emanuele M. Ventura}
\author[c]{Kenji Hasegawa,}
\author[a,d,e,f]{Kiyotomo Ichiki}

\affiliation[a]{Graduate School of Science, Nagoya University, Furocho, Chikusa-ku, Nagoya, Aichi, 464-8602, Japan}
\affiliation[b]{Research School of Astronomy and Astrophysics, Australian National University, Canberra, ACT 2611, Australia}
\affiliation[c]{National Institute of Technology, Suzuka College, Shirokocho, Suzuka, Mie 510-0294, Japan}
\affiliation[d]{Kobayashi-Maskawa Institute for the Origin of Particles and the Universe, Nagoya University, Furocho, Chikusa-ku, Nagoya, Aichi, 464-8602, Japan}
\affiliation[e]{Institute for Advanced Research, Nagoya University, Furocho, Chikusa-ku, Nagoya, Aichi 464-8602, Japan}
\affiliation[f]{National Astronomical Observatory of Japan, 2-21-1 Mitaka, Tokyo 181-8588, Japan}

\emailAdd{naruse.genki.v5@s.mail.nagoya-u.ac.jp}
\emailAdd{yuxiang.l.qin@gmail.com}
\emailAdd{lele.ventura@hotmail.it}
\emailAdd{hasegawa-k@mech.suzuka-ct.ac.jp}
\emailAdd{ichiki.kiyotomo.a9@f.mail.nagoya-u.ac.jp}

\abstract{
The 21 cm forest is a promising probe of small-scale structure in the high-redshift universe. However, the contribution of minihalos remains poorly understood because of the limited resolution of cosmological simulations and uncertainties in radiation feedback.
Here we investigate their contribution using a semi-analytical simulation \textit{MERAXES}, which resolves halo mass down to $\sim 2 \times 10^5 \Msunhinv$, combined with the analytical halo density and temperature profiles.  
In our fiducial model of the soft X-ray luminosity parameter per star formation rate $L_\mr{X<2keV}/\mr{SFR}\;[\mr{erg\;s^{-1}\;\Msun^{-1}\;\mr{yr}}] = 3.16 \times 10^{40}$, we find that absoptions with optical depth $\tau > \mathcal{O}(10^{-2})$ is dominated by minihalos, mostly those with masses of $\mathcal{O}(10^5)\Msunhinv$, whereas weaker absorption $\tau < \mathcal{O}(10^{-2})$ is dominated by the diffuse intergalactic medium (IGM).
At spectral resolution of $\Delta \nu \lesssim 100 \mr{Hz}$, the predicted abundance of strong absorption lines per unit redshift and logarithmic optical depth interval exceeds unity at the start of EoR. 
We also find that the distinction between minihalos and diffuse IGM absorption strongly depends on the IGM thermal history. 
}

\begin{document}

\maketitle
\flushbottom

\section{Introduction}
\label{sec:introI}

One of the most important open questions in modern cosmology is the nature of cold dark matter (CDM), particularly its clustering properties on small scales. 
While the $\Lambda$CDM model agrees well with cosmological observations at scales larger than $\sim 1\mr{Mpc}/h$ \cite{2020A&A...641A...1P}, the nature of dark matter remains less constrained on smaller scales, where different dark matter models can predict distinct halo abundances and internal structures. 
In the frame of the bottom-up structure formation scenario supported by the standard model, an interesting probe of small-scale structure is provided by "minihalos", which form before and around the first stage of the epoch of reionization (EoR) with the CDM mass from $\sim 10^4 \Msun$ to $\sim 10^7 \Msun$, and can retain neutral hydrogen without efficient atomic cooling.
At such a high-redshift universe, a substantial fraction of the hydrogen in the IGM and minihalos remains neutral. 
Thus, utilizing the signals from the 21 cm hyperfine transition is a favorable probe of these structures. 
Analogous to the Ly $\alpha$ forest, the 21 cm forest consists of absorption features imprinted on the spectra of bright, background radio sources or gamma-ray burst afterglows by intervening neutral hydrogen \cite{2002ApJ...577...22C, 2002ApJ...579....1F,2013MNRAS.428.1755C, 2015MNRAS.453..101C}. Dense HI structures, such as minihalos, can produce strong, narrow absorption features, making the 21 cm forest a promising probe of the physics at small scales in the high-redshift universe, such as the pre-reionization era \cite{2002ApJ...579....1F,2006MNRAS.370.1867F}.

For example, previous studies have shown that the 21 cm forest can probe the small-scale matter distribution and explore the DM models \cite{2014PhRvD..90h3003S, 2020PhRvD.101d3516S, 2020PhRvD.102b3522S, 2021JCAP...04..019K, 2023NatAs...7.1116S, 2025PhRvD.112f3513S, 2025PhRvD.112f3557S, 2026PhRvD.113h3525S}, neutrino masses \cite{2025PhLB..86239342S}, primordial black holes 
\cite{2023PASJ...75S..33V, 2026PhRvD.113d3531Z}, 
and DM-baryon relative velocity \cite{2023PhRvD.107l3520S}.
In addition, the 21 cm forest is sensitive to the thermal evolution of the intergalactic medium (IGM) \cite{2009ApJ...704.1396X, 2011MNRAS.410.2025X, 2012MNRAS.425.2988M, 2025MNRAS.537..364S}.
In particular, minihalos have been considered as potentially important contributors to the 21 cm forest \cite{2011MNRAS.417.1480M, 2023JCAP...03..017K, 2024JCAP...10..091N}.
Recent studies have also explored methods using one-dimensional power spectrum of the 21 cm forest \cite{2023NatAs...7.1116S,2025PhRvD.112f3513S} and non-Gaussian statistics to shed light on breaking the degeneracies between dark matter physics and the thermal history of the IGM \cite{2025PhRvD.112f3557S,2026PhRvD.113h3525S}. 

%
\begin{figure}[htbp]
\centering
\includegraphics[width=0.5\textwidth]{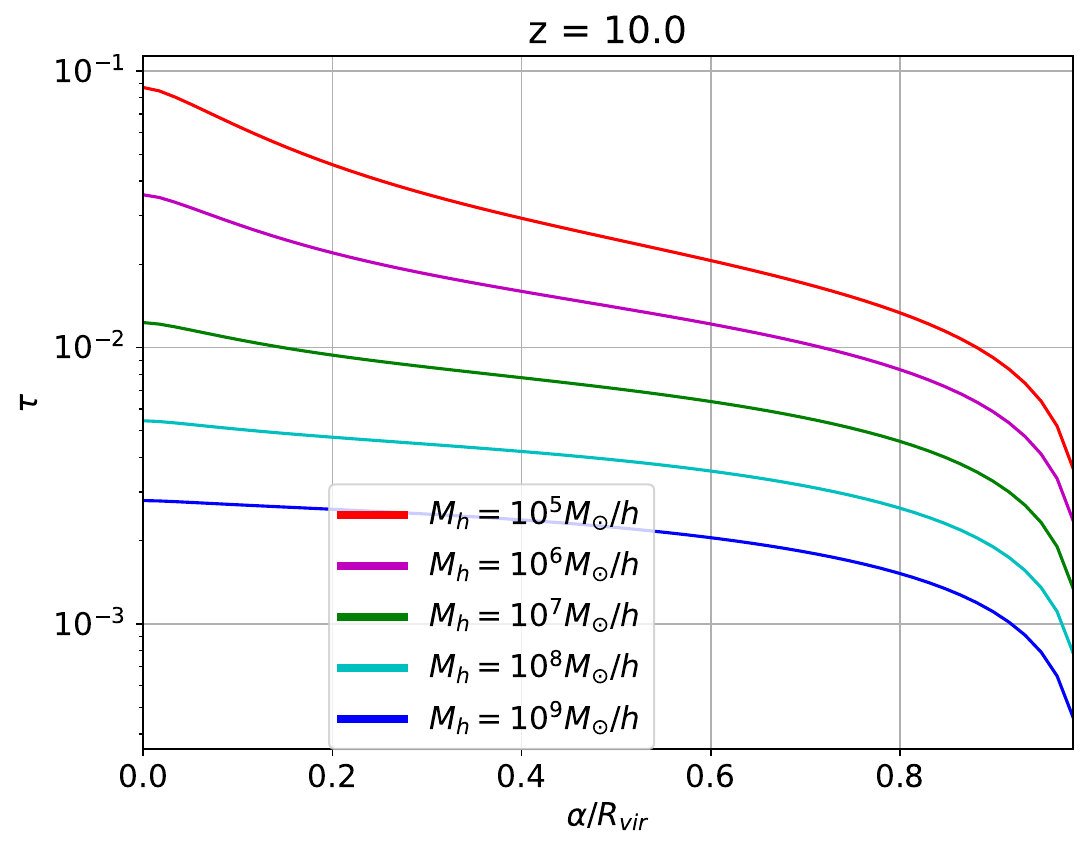}
\caption{
The 21 cm absorption optical depth $\tau$ through minihalos at $z = 10$, as a function of the impact parameter $\alpha$ from the halo center, normalized by the virial radius of the halos $R_\vir$.  Different lines correspond to different halo masses $M_h$. The optical depth is calculated following \cite{2023JCAP...03..017K, 2024JCAP...10..091N}, using the publicly available HAYASHI library \cite{2023JCAP...03..017K, pablo_villanueva_domingo_2022_7044255} (\url{https://github.com/PabloVD/HAYASHI}). 
}
\label{fig:EG_HaloTauRelation}
\end{figure}
The contribution of low-mass minihalos to the 21 cm forest depends on both the strength of their absorption and how often a sightline intersects them. Figure~\ref{fig:EG_HaloTauRelation} illustrates the first of these effects assuming the gas temperature follows minihalo's virial properties (Appendix~\ref{sec:haloanalyticprof}). It shows the optical depth $\tau$ through a minihalo at $z=10$ as a function of the impact parameter $\alpha$ in units of the virial radius $R_\vir$. We see that lower-mass minihalos produce stronger absorption at a given $\alpha/R_\vir$, primarily because of their lower virial temperatures. Although their smaller radii give them smaller geometric cross sections, their large abundance towards lower masses in CDM suggests minihalos could contribute appreciably to the incidence of strong absorption lines, provided they retain sufficient neutral gas.

Previous 21 cm forest calculations have often adopted a minimum halo mass of order \(10^{6}\,\Msun\) at \(z=10\), motivated by estimates of the Jeans or filtering mass \cite{2000ApJ...542..535G,2011MNRAS.410.2025X,2023NatAs...7.1116S}.
\footnote{
For example, one may claim that the minimum mass is enough to take into account some order of $10^6 \Msun$ around $z = 10$ considering the "filtering" mass (a time-averaged Jeans mass) with $f_\mr{X} (\lesssim 1)$ and CDM model \cite{2000ApJ...542..535G, 2011MNRAS.410.2025X, 2023NatAs...7.1116S}.  
However,
the definition of the Jeans mass and the filtering mass in those papers is different from
the definition of \cite{2001PhR...349..125B, 2011MNRAS.418..906T} by a factor of 8. 
The former definition of the Jeans mass $M_\mr{J}$ is $M_\mr{J} = 4\pi/3\bar{\rho}{\lambda_\mr{J}}^3$, the latter is $M_\mr{J} = 4\pi/3\bar{\rho}(\lambda_\mr{J}/2)^3$ with the same $\lambda_\mr{J}$ 
(and the filtering mass is also in the same manner). 
Even though an important role of the Jeans mass is the criterion set to consider the mass of the collapse object and the values themselves are not so strict, 
we take the latter definition in this study according to \cite{2001PhR...349..125B, 2011MNRAS.418..906T}. 
In this case, the filtering mass is $\sim 10^5 \Msun$ and the smaller halos whose mass is less than $10^6 \Msun$ should be important for the 21 cm optical depth around the first stage of EoR.
}
Whether halos below this mass contribute appreciably to the forest depends on how much neutral gas they can accrete and retain as reionisation proceeds. Heating of the surrounding IGM can suppress gas accretion onto shallow potential wells, while ionising radiation can drive the photoevaporation of gas already bound to a halo \cite{2004ApJ...600....1S,2004ApJ...610L...5S,2008ApJ...685...40W}. The Lyman--Werner background also affects minihalo evolution by dissociating molecular hydrogen and suppressing cooling and star formation \cite{2001ApJ...548..509M,2008ApJ...673...14O,2015MNRAS.448..568H}. Analytical studies have predicted absorption profiles for individual low-mass minihalos, but their gas prescriptions have typically omitted the effects of an evolving radiation background \cite{2020PhRvD.102b3522S,2020PhRvD.101d3516S,2023JCAP...03..017K}. 
On the other hand, cosmological simulations can follow this background across the IGM, but have generally lacked the mass resolution to resolve halos down to \(\sim10^{5}\,\Msunhinv\) \cite{2013MNRAS.428.1755C,2015MNRAS.453..101C,2023NatAs...7.1116S,2016MNRAS.455..962S}. 
Therefore, quantifying the absorption missed by a low-mass cutoff requires a model that captures the abundance and spatial distribution of these minihalos as well as the response of their gas to the local radiation field.
%

In this study, we investigate the minihalo contribution to the 21 cm optical depth in the forest, taking the effect of the background radiation into account through a semi-analytical prescription. 
For this purpose, we use the \textit{MERAXES} simulation data to obtain the spatial distribution of halos and their masses down to $M_\mr{h} \sim 2 \times 10^{5} \Msunhinv$, together with the thermal and ionization histories of the IGM. We then assign the analytical gas density profile and temperature profile to individual halos in the simulation results and calculate the 21 cm forest spectra along the lines of sight. By comparing contributions from minihalos and the IGM, we quantify the impact of low-mass minihalos on the number and distribution of absorption-line optical depths, and investigate their dependence on frequency resolution, redshift, and the X-ray heating history.
This study focuses on characterizing the absorption signatures of minihalos, leaving constraints on cosmological and astrophysical parameters for future work.

The composition of our paper is as follows. 
Section \ref{subsec:meraxesinfo} introduces the simulation we use, and we explain how to estimate the 21 cm forest, including the minihalo contribution from the simulation data, in Section \ref{subsec:haloassess} and Section \ref{subsec:tauformula}. 
Then, in Section \ref{sec:results}, we present our results and estimate how minihalo absorption lines will appear in the 21 cm forest.
Section \ref{sec: discussion} is devoted to the discussion, and finally, we summarize this study in Section \ref{sec: summary}.
Throughout this paper, we assume the \lcdm model with the following cosmological parameters: $h = 0.6751$, $\Omega_\mr{m} = 0.3121$, $\Omega_\mr{b} = 0.0490$, $\Omega_\Lambda = 0.6879$. These values are based on the Planck 2015 result \cite{2016A&A...594A..13P} and are the default fiducial values set by the simulation \textit{MERAXES}, which will be introduced in the next section.

\section{Methodology}
\label{sec: methodology}

\subsection{Simulation Setup}
\label{subsec:meraxesinfo}

To investigate the low-mass minihalo contribution to the 21 cm forest, we use the semi-analytic galaxy formation model \textit{MERAXES} \cite{2016MNRAS.462..250M, 2017MNRAS.472.2009Q, 2019MNRAS.489.1357Q, 2024MNRAS.529..628V}, coupled to the high-resolution GenesisL10\_N2048\footnote{The \textit{MERAXES} model is publicly available on GitHub: \url{https://github.com/meraxes-devs/meraxes} while the merger trees are available on Zenodo: \url{https://zenodo.org/records/10608236}} \textit{N}-body simulation \cite{2024MNRAS.529..628V}. 
\textit{MERAXES} is designed to investigate the galaxy formation history during the EoR, incorporating a modified version of \textit{21cmFAST} \cite{2011MNRAS.411..955M}. The GenesisL10\_N2048 simulates the evolution of structure formation, providing dark matter halo catalogs and merger trees around the beginning of the reionization era.

\begin{figure}[htbp]
\centering
\includegraphics[width=0.5\textwidth]{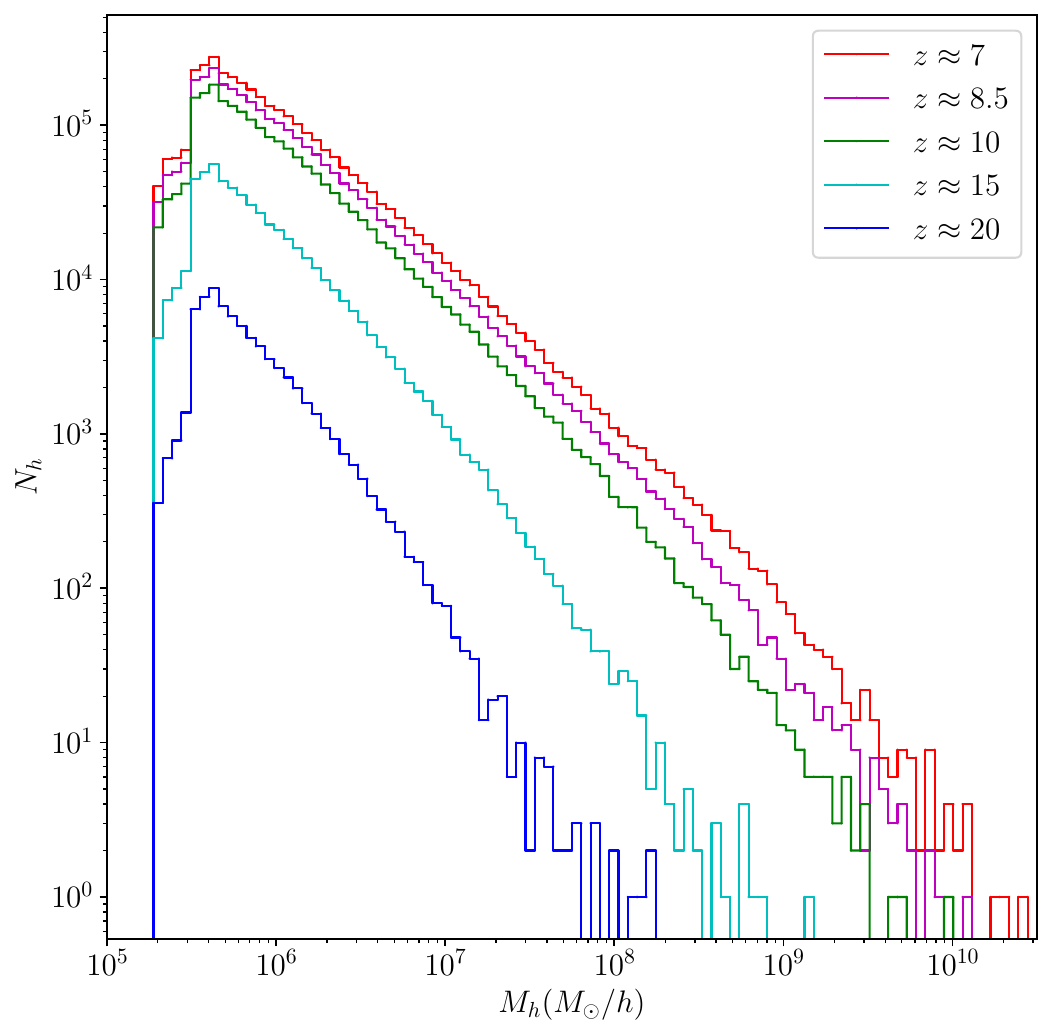}
\caption{
Halo mass distributions in the Genesis L10\_N2048 simulation at different redshifts, as indicated in the figure.
The vertical axis shows the number of halos $N_\mr{h}$ in each mass bin. As mentioned in the main text, \textit{MERAXES} can trace the minimum halo mass down to $M_\mr{h} \sim 2 \times 10^5 \Msunhinv$. 
}
\label{fig:halocount}
\end{figure}

The simulation contains $2048^3$ DM particles whose mass is $\sim 9.935 \times 10^3 \Msunhinv$ in a comoving box of size $(10 \mr{cMpc}/h)^3$. We include halos down to a mass of approximately $\sim 2 \times 10^5 \Msunhinv$, as shown in Figure \ref{fig:halocount}. 
The number of grids is 128 on a side, giving a length of approximately $\sim 78.1h^{-1} \mr{ckpc}$, where the physical properties on the field such as density fluctuation, temperature, and radiation background, and so on, are given. 
We adopt the fiducial astrophysical parameters of Ventura et al. (2024) \cite{2024MNRAS.529..628V}, with one change to the escape fraction of ionizing photons: we reduces its normalization from $f_\mr{esc}^0=0.06$ to $0.04$ while retaining the redshift dependence $f_\mr{esc} = f_\mr{esc}^0 \left[ (1+z)/6\right]^{0.5}$ \cite{2017MNRAS.472.2009Q, 2023MNRAS.520.3368B}. The left panel of Figure~\ref{fig:xHI_T_Frac_history:fid} shows the resulting evolution of the volume-averaged neutral hydrogen fraction, $x_\mr{HI}$.
To investigate the effect of the X-ray heating, we vary the soft-band X-ray luminosity per star formation rate (SFR) $L_\mr{X<2keV}/\mr{SFR}\;[\mr{erg\;s^{-1}\;\Msun^{-1}\;\mr{yr}}]$ with $3.16 \times 10^{40}$ as our fiducial value \cite{2023MNRAS.520.3368B, 2025MNRAS.540..483V}. For this value, the middle panel of Figure~\ref{fig:xHI_T_Frac_history:fid} shows the evolution of the volume-averaged IGM kinetic temperature $T_\mr{k}$ and spin temperature $T_\mr{s}$ alongside the CMB temperature $T_\mr{cmb}$.
Other X-ray heating models are introduced in Section \ref{subsec: resultwithIGM}. 
Further details of the \textit{MERAXES} simulation can be found in \cite{2016MNRAS.462..250M, 2017MNRAS.472.2009Q, 2019MNRAS.489.1357Q, 2020MNRAS.495..123Q, 2023MNRAS.520.3368B, 2024MNRAS.529..628V}. 

\subsection{Minihalo classification and HI gas abundance}
\label{subsec:haloassess}

The GenesisL10\_N2048 simulation provides the masses and spatial distributions of dark matter halos. 
To calculate their contributions to the 21 cm forest, we assign analytical density and temperature profiles to individual halos.
We model the gas as a spherically symmetric, isothermal atmosphere in hydrostatic equilibrium with a Navarro, Frenk \& White (NFW) DM potential \cite{2000ApJ...540...39A, 1997ApJ...490..493N}. The gas temperature is assumed to be equal to the virial temperature of each halo. Details of the adopted profile are given in appendix \ref{sec:haloanalyticprof}.

\begin{figure}[htbp]
\centering
\includegraphics[width=1.0\textwidth]{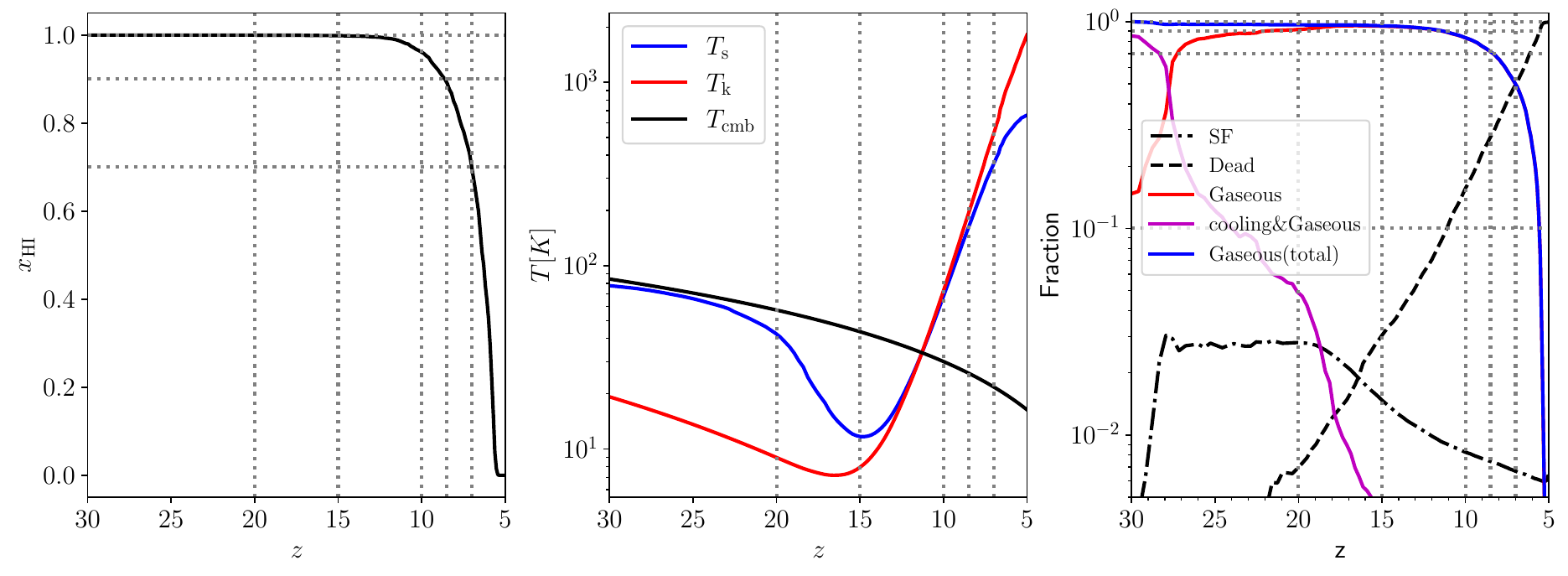}
\caption{Redshift evolution of the ionization and thermal states of the IGM and the minihalo populations in the fiducial model, with $L_\mr{X<2keV}/\mr{SFR}\;[\mr{erg\;s^{-1}\;\Msun^{-1}\;\mr{yr}}] = 3.16 \times 10^{40}$.
The left panel shows the history of the volume-averaged HI fraction ($x_\mr{HI}$). 
The middle panel shows the history of the volume-averaged spin temperature ($T_\mr{s}$, blue) and gas temperature ($T_\mr{k}$, red) with the CMB temperature ($T_\mr{cmb}$, black). 
The right panel shows the history of the number fractions of various halos: star-forming (SF), gaseous, and gas-depleted (Dead). 
The "cooling \& Gaseous" category in magenta represents halos that satisfy both the gas-retention and molecular-cooling criteria but contain no stars in the simulation. These halos are included in the gaseous population, and "Gaseous (total)" in blue denotes the sum of the two gaseous categories.
}
\label{fig:xHI_T_Frac_history:fid}
\end{figure}

The gas content of minihalos is affected by radiative feedback associated with star formation. 
Molecular hydrogen cooling enables star formation in sufficiently massive minihalos, while the Lyman-Werner background suppresses this cooling process. Ionizing UV radiation can heat and photoevaporate gas from low-mass halos to reduce their gas content. To approximately account for these effects, 
we classify halos into three categories: star-forming (SF), gaseous, and gas-depleted (hereafter referred to as Dead) minihalos. 
The classification is based on the minimum halo mass for star formation and the characteristic mass for gas depletion, as introduced below. 
The right panel of Figure \ref{fig:xHI_T_Frac_history:fid} shows the ratio of these types of halos in the simulation.

We first determine whether a halo can form stars using the minimum mass criterion for molecular hydrogen cooling \cite{2015MNRAS.453.4456V,2024MNRAS.529..628V}, 
\begin{equation}
    \label{equation:m_mc}
    M_\mr{critMC} = M_\mr{cool, H_2} \left[1+ 6.96\left( 4\pi J_\mr{LW}(z) \right)^{0.47}\right]~.
\end{equation}
Here $M_\mr{cool,H_2}$ is the minimum halo mass required for efficient molecular hydrogen cooling, including the effect of the DM-baryon streaming velocities \cite{2010PhRvD..82h3520T}. 
The quantity $J_\mr{LW}$ is the Lyman-Werner background flux in the unit of $10^{-21} \mr{erg \; s^{-1} \; cm^{-2} \; Hz^{-1} \; sr^{-1}}$. 
We next account for the suppression of gas accretion and retention by the ionizing UV background. Following Sobacchi \& Mesinger (2013) \cite{2013MNRAS.432L..51S}, 
we adopt a characteristic halo mass $M_{\rm filt}$, at which the baryon fraction is reduced to half of the cosmic mean value. This mass is given by \cite{2016MNRAS.462..250M, 2013MNRAS.432L..51S, 2013MNRAS.432.3340S}, 
\begin{equation}
    \label{equation:mfilt}
    M_\mr{filt} = M_0 J_{21}^a \left( \frac{1+z}{10} \right)^b \left[ 1 - \left( \frac{1+z}{1+z_\mr{ion}} \right)^c \right]^d, 
\end{equation}
where $\left(M_0, a,b,c,d \right) = \left( 2.8 \times 10^9 \Msun, 0.17, -2.1, 2.0, 2.5\right)$, $z_\mr{ion}$ is the redshift when the halo is first exposed to an ionizing UV background, and $J_\mr{21}$ represents the local UV background intensity at $\nu=\nu_{\rm H}$, defined by 
\begin{equation}
    J(\nu) = J_{21} \left(\frac{\nu}{\nu_\mr{H}}\right)^{-\alpha} \times 10^{-21} \mr{erg \; s^{-1} \; Hz^{-1} \; cm^{-2}\;sr^{-1}}
\end{equation}
where $\nu_\mr{H}$ is the Lyman limit frequency and $\alpha = 5.0$ corresponds to a stellar-driven UV spectrum \cite{1996ApJ...465..608T}.
The baryon fraction of a halo is then reduced by a factor $f_\mr{mod} = 2^{-M_\mr{filt}/M_\mr{vir}}$, following the implementation in MERAXES
\footnote{Past studies using the hydrodynamical simulations showed that the filtering mass is agreed well with this critical mass that is defined as the mass whose baryon components remains half of the original gas mass fraction ($\Omega_\mr{b}/\Omega_\mr{m} =0.1571$ in this study) \cite{2000ApJ...542..535G, 2008MNRAS.390..920O, 2009MNRAS.399..369N}.}.

Therefore, the classification of halos from simulation data according to their masses and their HI gas abundance are like below:
\begin{itemize}
    \item Star-forming (SF) halos:  
    Halos who have a nonzero stellar mass from the simulation result.
    Their baryon fractions are given by ${f_\mathrm{b}} = {\Omega_b}/{\Omega_m} \times f_\mr{mod}$, and their neutral hydrogen fractions $x_\mr{HI}$ are estimated from the corresponding simulation grid cells. 
    This treatment of $x_\mr{HI}$ for SF halos is approximate, since their internal ionization states may differ from the grid-averaged values depending on the mass of the halo, the mass of the star, and the elapsed time from the star formation, etc. 
    However, as shown in the right panel of Figure \ref{fig:xHI_T_Frac_history:fid}, SF halos constitute only a small fraction of the halo population at high redshift in this study and 
    the gaseous minihalos dominate.
    Therefore, we expect the impact of this approximation on the 21 cm forest to be limited.
    \item Dead halos: 
    We classify starless halos with $M_\vir < M_\mr{filt}$ as Dead. For these halos, we neglect any residual gas and set $f_\mathrm{b}=0$ and $x_\mr{HI}=0$. When $M_\mr{filt} \geq M_\mr{critMC}$, this class also includes starless halos with $M_\mr{critMC}\leq M_\vir\leq M_\mr{filt}$. They meet the mass criterion for molecular cooling but fall at or below the threshold we adopt for gas retention.
    \item Gaseous halos: 
    We classify starless halos with $M_\vir \geq M_\mr{filt}$ as gaseous. We assign them a baryon fraction $f_\mathrm{b}=(\Omega_\mathrm{b}/\Omega_\mathrm{m})f_\mr{mod}$ and assume that their hydrogen is neutral ($x_\mr{HI}=1$). Within this population, halos with $M_\vir<M_\mr{critMC}$ retain gas but lie below the molecular-cooling threshold. Those with $M_\vir\geq M_\mr{critMC}$ also meet the cooling criterion, although star formation in \textit{MERAXES} additionally requires a sufficiently large gas reservoir. We label this latter subset ``cooling \& Gaseous'' in the right panel of Figure~\ref{fig:xHI_T_Frac_history:fid} and include it in the total gaseous fraction.
\end{itemize}

\subsection{Calculation of the 21 cm optical depth in the spectrum}
\label{subsec:tauformula}

Using the halo catalogs and the assigned gas profiles, we generate 21 cm forest spectra along $128^2$ lines of sight (LOSs) through the simulation volume.
The total optical depth at each frequency is calculated by combining the contributions from minihalos and the diffuse IGM.

\begin{figure}[htbp]
\centering
\includegraphics[width=0.7\textwidth]{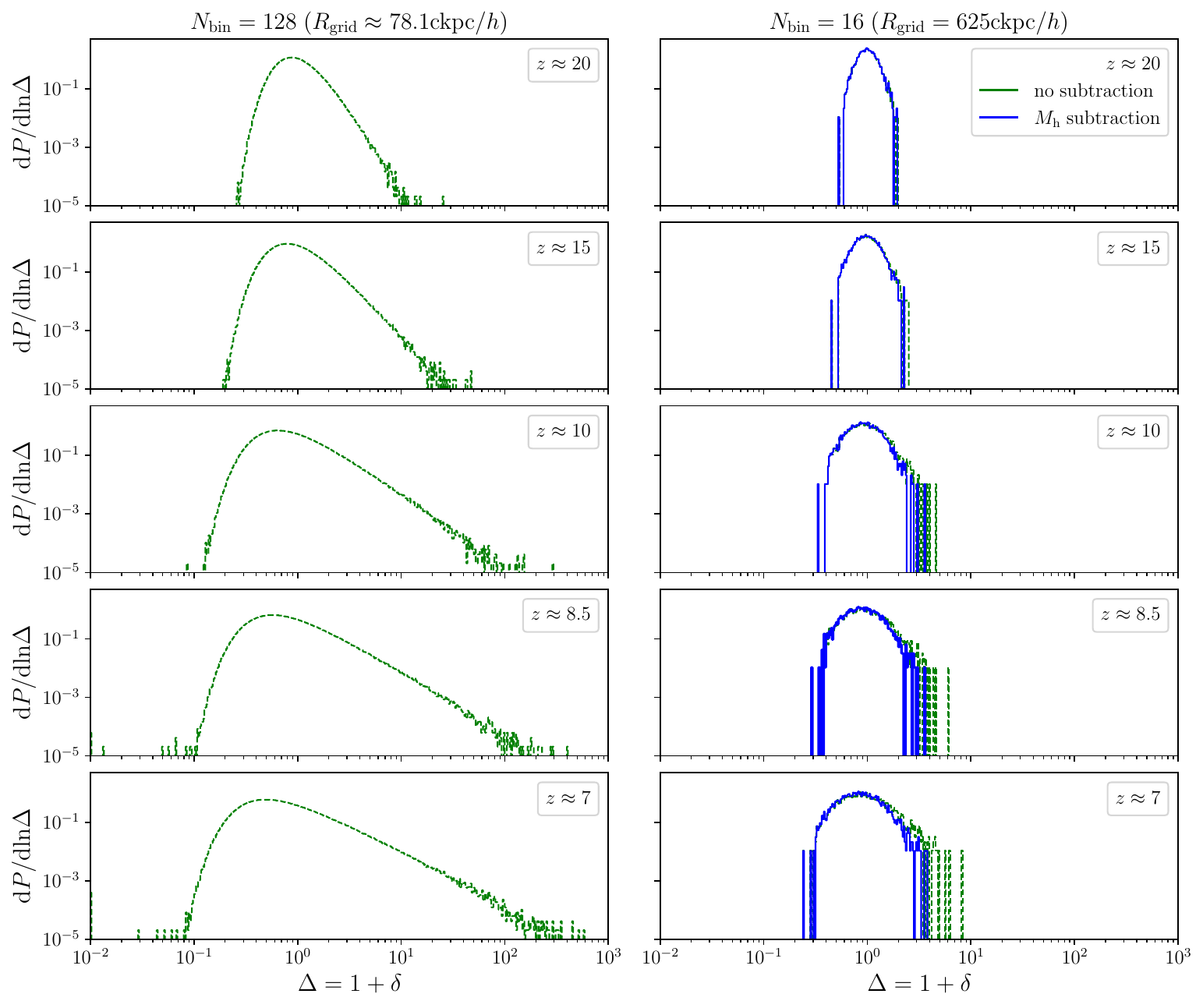}
\caption{
The PDF of the density fluctuation $\Delta = 1 + \delta$. 
The left panels show the simulation data gotten by the number of the grids on a side is 128. 
The right panels are by the smoothing with the $16$ grids on a side. 
The blue solid lines represent the results by subtracting the halos mass from the mass in the smoothed grids, whereas the green dashed lines do not subtracting results. 
}
\label{fig:overdensity_history}
\end{figure}

To model the IGM, we construct a smoothed density field by reducing the spatial resolution from $128$ to $16$ grids, corresponding to a cell size from ($\sim 78.1 \mr{ckpc}/h$) to ($625 \mr{ckpc}/h$). We then subtract the mass associated with halos from each cell to remove their contributions to the diffuse IGM, for which we require $\Delta = 1 + \delta \lesssim 10$ is satisfied. The distribution of $\Delta$ of the diffuse IGM is shown as blue lines in Figure \ref{fig:overdensity_history}. 
Additionally, the temperature is also smoothed onto the same grid. 

After obtaining physical properties of the diffuse IGM, the 21 cm optical depth by the diffuse IGM at the frequency $\nu_0$ on the LOS (corresponding to $z_0$ and $\nu_0(1+z_0) = \nu_{21}$) is calculated by \cite{1959ApJ...129..525F, 1997ApJ...475..429M, 2002ApJ...579....1F,2006PhR...433..181F, 2013MNRAS.428.1755C, 2015MNRAS.453..101C}
\begin{align}
    \label{tauofIGM:eq}
    \tau{\left(\nu_0\right)}
    &= \frac{3 h_p c^3 A_{10}}{32 \pi k_B \nu_{21}^2 } 
    \frac{n_\mathrm{HI}(z_0)}{T_\mr{S}(z_0)\mathrm{d}v(z_0)/\mathrm{d}r}\\
    &\approx 6.967 \times 10^{-3} \left( 1+\delta(z_0) \right) \left( 1+z_0\right)^{3/2} \frac{x_\mr{HI}(z_0)}{T_\mr{S}(z_0)} \left[ \frac{H(z_0)}{\mathrm{d}v(z_0)/\mathrm{d}r} \right], \notag
\end{align}
where $h_\mr{p}$ is the Planck constant, $c$ is the speed of light, $A_{10}$ is the Einstein coefficient, $k_\mr{B}$ is the Boltzmann constant, $\nu_{21}$ is the frequency corresponding to 21 cm line, $\mathrm{d}v/\mathrm{d}r$ is the velocity gradient along the LOS, $T_\mr{S}$ is the spin temperature, and $n_\mr{HI} = \bar{n}_\mr{H}x_\mr{HI}(1+\delta)$ is the HI number density.
In our simulation, we neglect peculiar velocity gradients in the diffuse IGM and assume ${\rm d}v(z_0)/{\rm d}r = H(z_0)$. On the scales over which thermal broadening contributes to the absorption at a given frequency, we treat the gas properties of the smoothed IGM as uniform. Within a minihalo, the neutral hydrogen density and spin temperature vary along the sightline, and the local kinetic temperature sets the thermal width of the line. We therefore integrate the absorption through each intersected halo using its radial gas profiles (see Appendix~\ref{sec:tauhalooverestimate}).

\begin{figure}[htbp]
\centering
\includegraphics[height=0.3\textheight]{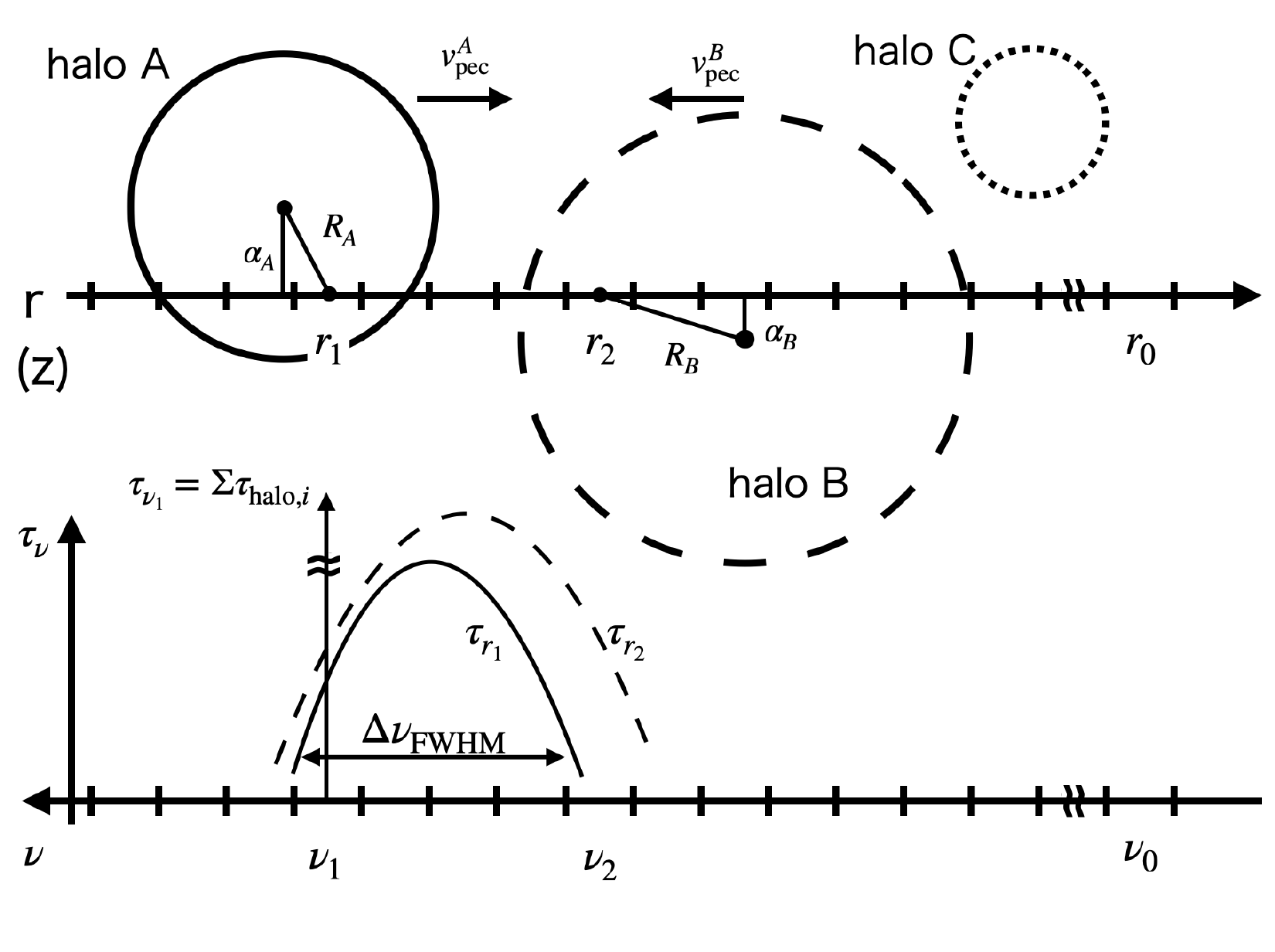}
\caption{
Schematic illustration of the 21 cm optical-depth calculation along a line of sight intersecting multiple minihalos. The upper panel shows three halos: halos A and B intersect the LOS, while halo C does not. The lower panel illustrates the absorption profiles associated with gas in halos A and B. The profiles are thermally broadened and shifted in frequency by the peculiar velocities of the individual halos. Consequently, gas at different positions along the LOS can contribute to the optical depth at the same observed frequency. The total optical depth is calculated by summing the contributions from all intersected halos.
For the pixel where any halo does not contribute, such as $r_0$, the optical depth is given by the diffuse IGM and can be calculated using equation (\ref{tauofIGM:eq}). 
}
\label{fig:schematicview}
\end{figure}

When a LOS crosses a minihalo at redshift $z_1$ with an impact parameter $\alpha$ from the center (see Figure \ref{fig:schematicview}), we calculate the optical depth by the halo at frequency $\nu_1= \nu_{21}/(1+z_1)$ as  \cite{2002ApJ...579....1F}
\begin{equation}
    \label{tauofmh:eq}
    \tau\left(\nu_1\right) 
    = \frac{3 h_\mathrm{p} c^3 A_{10}}{32 \pi k_\mathrm{B} \nu_{21}^2}\int
    ^{\infty}_{-\infty}
    \mathrm{d}r'
    \frac{n_\mathrm{HI}\left( R \right)}{T_\mathrm{s}\left( R \right) \sqrt{\pi} b(R)}\exp{\left(-\frac{v^2(\nu',z_1)}{b^2(R)}\right)}, 
\end{equation}
Here $r'$  is the LOS coordinate measured from the point of closest approach to the halo center, and  $R=\sqrt{\alpha^2+r'^2}$ where $\alpha$ is the impact parameter and $R$ is the radial distance from the halo center. 
The frequency $\nu'$ corresponds to the cosmologically-redshifted frequency of the 21cm line at position $r'$ along the LOS.
The quantities $n_\mr{HI}(R)$, $T_\mr{S}(R)$, $T_\mr{K}(R)$ inside the halo follow the analytical gas profiles described in Sec.~\ref{subsec:haloassess}. Thermal Doppler broadening is characterized by $b^2(R) = 2 k_\mr{B} T_\mr{k}(R)/m_\mr{p}$, and 
$v(\nu',z_1)  = c(\nu' - \nu_{1})/\nu_{1} + v_\mr{pec}(R)$
with the halo peculiar velocity $v_\mr{pec}$ along the LOS whose positive sign is taken to the direction going away from us.
{We adopt the peculiar velocity of each halo from the simulation and neglect internal bulk motions and nonthermal velocity dispersion. When multiple halos contribute at the same observed frequency, their optical depths are summed, accounting for their individual peculiar velocities.}

Figure \ref{fig:schematicview} shows our method for calculating the optical depth by halo using equation (\ref{tauofmh:eq}). 
When the LOS passes through halo A, the physical properties at the pixel inside the halo, such as $r_1$ (corresponds to $\nu_1$), are specified according to the profiles of the halo A with $\alpha_A$ and $R_A$. 
Additionally, the 21cm optical depth distribution along the sightline experiences Doppler broadening  due to the thermal velocity of HI. The peculiar velocity of the halo shifts the central frequency of the profile, causing a redshift in the optical depth profile by HI at $r_1$ as shown in the Figure. 
This line broadening, combined with the shift caused by peculiar velocity, leads to overlapping contributions to the optical depth at $\nu_1$ from other halos that the LOS intersects. 

For example, if halo B in the Figure has a peculiar velocity that results in a blueshift of its profile, the optical depth profile for HI at $r_2$ (corresponding to $\nu_2$), which is defined by the physical properties of halo B with parameters $\alpha_B$ and $R_B$, may also affect the optical depth at $\nu_1$. 
Therefore, we must take a sufficiently wide integration range in equatioin (\ref{tauofmh:eq}) to accurately calculate the optical depth at the pixel. 
In our calculations, we specify that the width of the line broadening is represented by the full width at half maximum (FWHM): $\Delta \nu_\mr{FWHM} = 2\sqrt{\ln 2}\; \nu_{21}b/c$ in the rest frame, which helps reduce the computational cost of the integration. 

There are two important points to consider regarding the frequency resolution. 
First, to properly account for the contribution of the halos, we need to calculate the optical depth with a frequency resolution that is finer than the scale of the halos. 
If the properties of the pixel are based on the halo profiles but the pixel length exceeds the diameter of the halo, the calculation result will be artificially overestimated.

In this study, we focus on frequency resolutions of $10\mr{kHz}$, $5.4\mr{kHz}$, $1\mr{kHz}$, $100\mr{Hz}$, $54\mr{Hz}$, $14\mr{Hz}$. 
The $5.4 \mr{kHz}$ resolution is the standard for the Square Kilometre Array (SKA) \cite{2019arXiv191212699B}
. 
We also consider results down to $14\mr{Hz}$ because there are some SKA projects to zoom up to that resolution \cite{2026aask.bookE..25C} \footnote{For more information, refer to the "Design Baseline Description" in the SKA Observatory HP \url{https://www.skao.int/en/science-users/118/ska-telescope-specifications}}, and thus our findings can serve as a reference for future studies. 
Therefore, we take the highest resolution of the pixel for the calculations as $\Delta \nu_\mr{hr} < 14 \mr{Hz}$, which is sufficiently smaller than the scale of a halo with a mass of $10^5 \Msunhinv$. For example, the virial radius $R_\mr{vir}(M_\mr{h} = 10^5 \Msunhinv) \approx 1.553 \times 10^2 \mr{pc}$ at $z \approx 10$ corresponds to a frequency change of approximately $\Delta \nu \approx 1.164 \times 10^2 \mr{Hz}$. 
Second, the equation (\ref{tauofIGM:eq}) is generally applicable only in cases where $n_\mr{HI}$, $T_\mr{S}$, $T_\mr{k}$ and $v_\mr{pec}$ can be treated as constants throughout the integration range in equation (\ref{tauofmh:eq}) \cite{2002ApJ...579....1F}. 
If equation (\ref{tauofIGM:eq}) is applied to pixels located within halos or in regions where the FWHM of multiple halos overlap, and the physical properties are not uniform in the integration range, the resulting calculations may be inaccurately distorted (see Appendix \ref{sec:tauhalooverestimate}). 
Hence, we note that these $n_\mr{HI}$ and $T_\mr{S}$ in equation (\ref{tauofIGM:eq}) should be the properties of the diffuse IGM\footnote{
In this regard, if all the fine pixels have different physical properties, we should use equation (\ref{tauofmh:eq}), instead of (\ref{tauofIGM:eq}) to calculate the optical depth of the diffuse IGM. 
However, by our definition of "diffuse" IGM, the properties at these pixels are constants in a smoothed grid, whose size is larger than the integration length of the equation (\ref{tauofmh:eq}), which corresponds to the FWHM of the diffuse IGM in this case. 
Therefore, we can use equation (\ref{tauofIGM:eq}) to determine the IGM optical depth in this study, even when the pixels are as fine as structures within a halo. 
}. 

\begin{figure}[htbp]
\centering
\includegraphics[height=0.4\textheight]{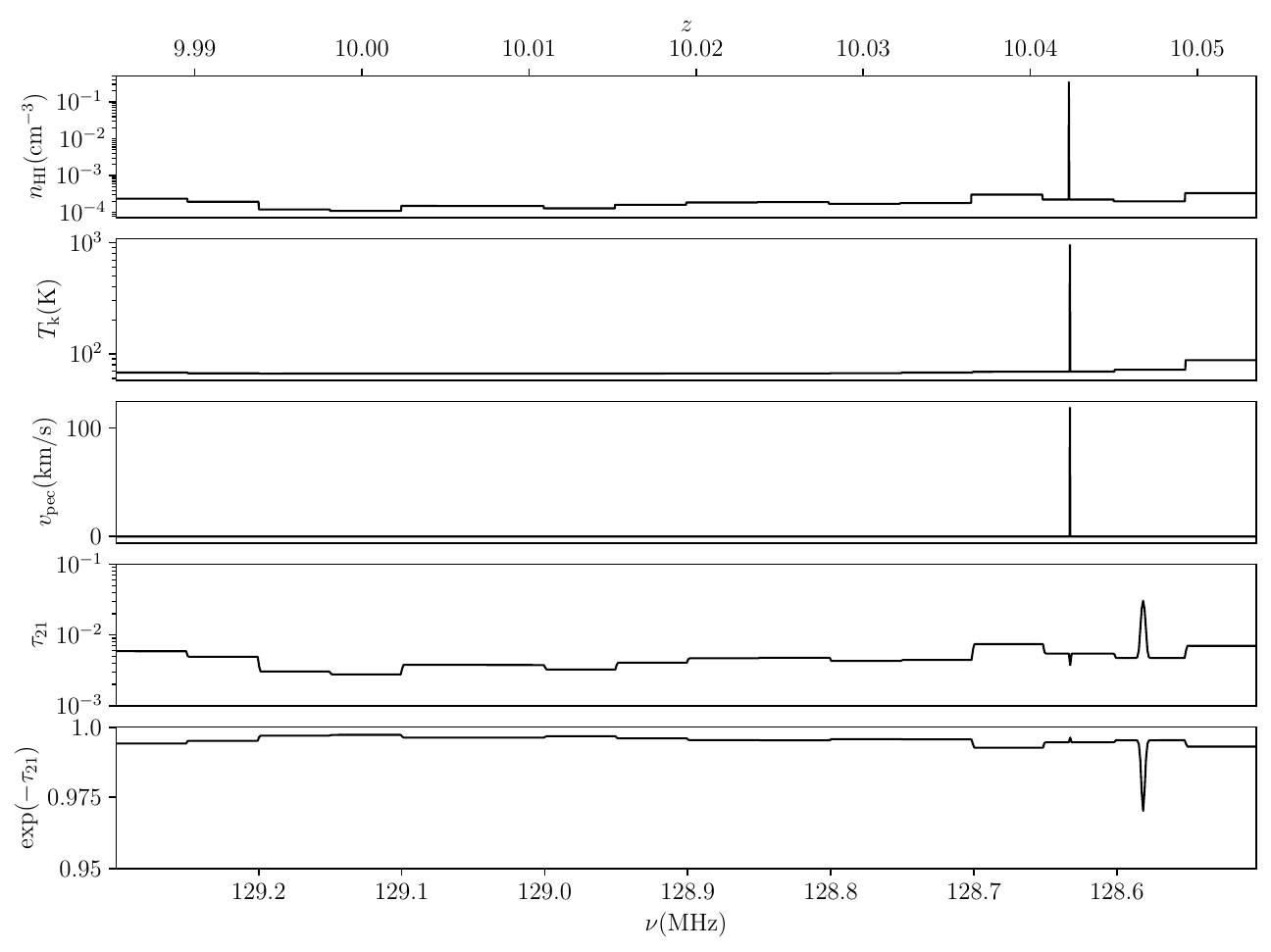}
\caption{
An example of the spectrum along a LOS at $z \sim 10$. From top to bottom, the panels illustrate the HI number density, the gas temperature, the peculiar velocity along the LOS, the 21 cm optical depth, and the transmitted flux, respectively. 
The last two measurements were obtained with a smoothing width of $1 \mr{kHz}$.
We note that we neglect any noise in this study. 
}
\label{fig:spectrumex2}
\end{figure}

Thus, with equations (\ref{tauofIGM:eq}) and (\ref{tauofmh:eq}), we first calculate the optical depths at the highest resolution, $\tau_\mr{hr}$, and then we convolve these results to obtain the optical depths at the lower resolution, $\tau_\mr{lr}$, we aim for:
\begin{equation}
        \label{tauconv:eq}
    \tau_\mr{lr}(\nu) = \frac{\int \mathcal{W}(\nu - \nu'_\mr{hr}) \tau_\mr{hr}(\nu'_\mr{hr}) \mr{d}\nu'_\mr{hr}}{\int \mathcal{W}(\nu - \nu'_\mr{hr}) \mr{d}\nu'_\mr{hr}},
\end{equation}
where
\begin{equation}
    \mathcal{W}(\nu - \nu'_\mr{hr}) = 
    \begin{cases}
        1 & \text{$\left(|\nu - \nu'_\mr{hr}|<\Delta \nu_\mr{lr}/2\right)$}\\
        0 & \text{otherwise}
    \end{cases}
    ,
\end{equation}
assuming $\tau \ll 1$.
Figure \ref{fig:spectrumex2} shows an example of the properties along the LOS. 
The panels, from top to bottom, represent the HI number density, the gas temperature, the peculiar velocity along the LOS (with positive values indicating redshifting away from us), the optical depth, and the transmitted flux $\exp{(-\tau_{21})}$, respectively. 
The spectra of the optical depth and the transmitted flux are smoothed with a resolution of $\Delta \nu \approx 1 \mr{kHz}$.
The sharp peak around $\nu \sim 128.64 \mr{MHz}$ (corresponding to $z \sim 10.042$) in the three panels from the top indicates the presence of a minihalo, while the other uneven parts represent the diffuse IGM.

The peculiar velocity of the halo causes the peak of the optical depth to be redshifted, and resulting in its appearance at a lower frequency compared to the peaks shown in the top three panels. 
Small artificial dips around $\nu \sim 128.64 \mr{MHz}$ in the optical depth and in the transmitted flux are remnants of this shift caused by the peculiar velocity. 
These dips appear near the original position of the halo absorption feature because the halo contribution is Doppler-shifted, whereas peculiar velocities are neglected for the diffuse IGM. They are numerical artifacts arising from the different treatments of the velocity fields and should not be interpreted as physical absorption features.

\section{Results}
\label{sec:results}

After estimating the $128^2$ 21 cm forest spectra, we calculate the abundance of the absorption lines per logarithmic optical depth per redshift, $\mrd^2 N/\mrd \ln \tau \mrd z$, and estimate the halo contribution. 

\subsection{Resolution dependence and halo mass dependence}
\label{subsec: mhtau}

\begin{figure}[htbp]
\centering
\includegraphics[width=1.0\textwidth]{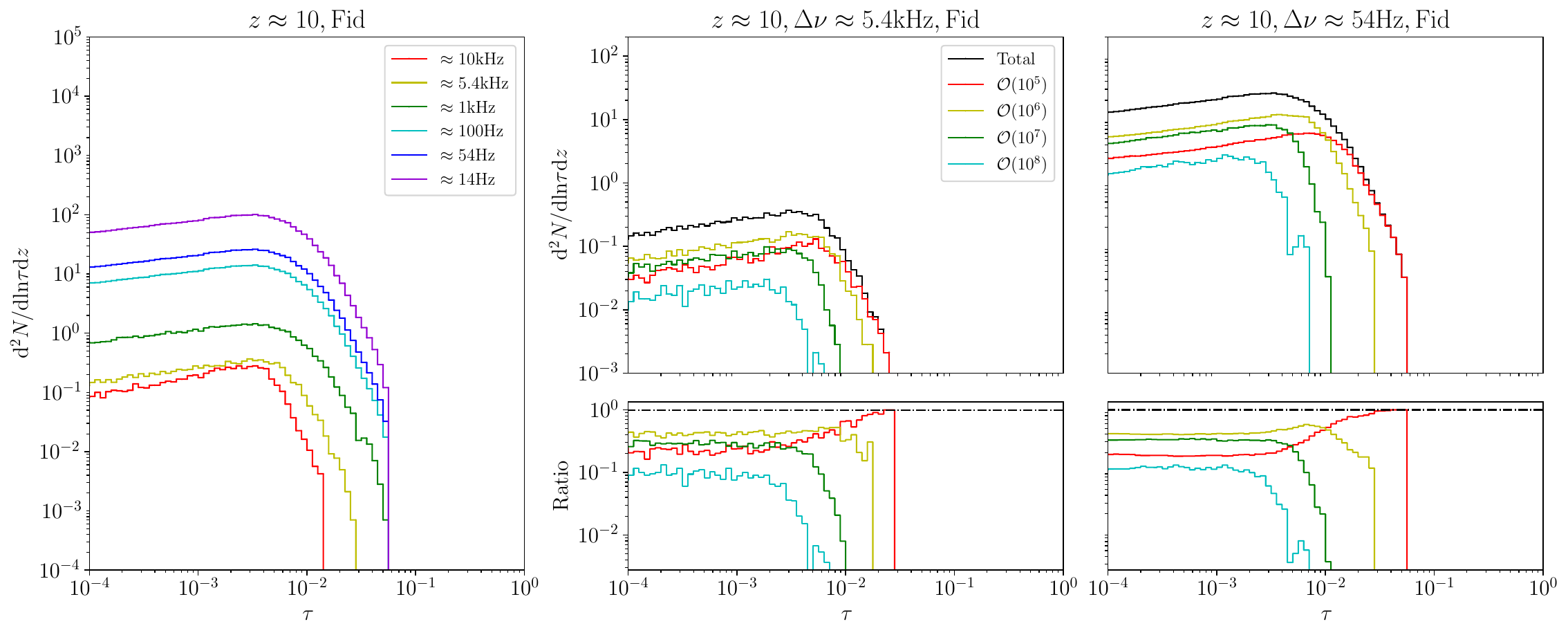}
\caption{
$\mrd^2 N/\mrd \ln \tau \mrd z$ at $z \approx 10.0$ in the fiducial model. 
This optical-depth abundance is attributed solely to minihalos. 
In the left panel, the abundance of the optical depth is displayed for the differences in $\Delta \nu$, with each color representing a specific range.
The middle and right panels show the dependence of optical depth on the minihalo mass.
The upper two panels show the abundance of the optical depth for minihalos across different mass ranges. For example, a notation of $\mathcal{O}(10^6)$ means a mass range of $10^6 \Msunhinv \leq M_\mr{vir} < 10^7 \Msunhinv$. 
The lower two panels show the ratio of these abundances to the total abundance.
The middle panels present results for $\Delta \nu \approx 5.4 \mr{kHz}$, while the right panels correspond to $\Delta \nu \approx 54 \mr{Hz}$. 
}
\label{fig:dNdlntdz:MHFidz10:res}
\end{figure}

Our main results are shown in Figure \ref{fig:dNdlntdz:MHFidz10:res}. 
The left panel shows the abundance at various frequency resolutions $\Delta \nu$.
A finer resolution leads to 
a larger maximum value of the optical depth and an increased number of greater optical depth values. 
This is because finer resolutions can resolve lower-mass minihalos, which have a larger optical depth. Additionally, within the same halos, higher resolution can resolve the inner regions, where the optical depth is greater toward the halo center, as shown in Figure \ref{fig:EG_HaloTauRelation}. 
For example, at $z \approx 10$, for a halo mass of $M_\mr{h} = 10^5 \Msunhinv$, the virial radius is approximately $R_\mr{vir}(M_\mr{h} = 10^5 \Msunhinv) \approx 1.553 \times 10^2 \mr{pc}$, or roughly $\sim 92.1 \mathrm{Hz}$. In comparison, for a larger halo with a mass of $M_\mr{h} = 10^6 \Msunhinv$, the virial radius is $R_\mr{vir}(M_\mr{h} = 10^6 \Msunhinv) \approx 3.346 \times 10^2~\mr{pc}$, equivalent to $\sim 199~\mathrm{Hz}$. Therefore, a resolution of $\Delta \nu \lesssim 100~\mr{Hz}$ can resolve the small halo enough and the abundance of $\tau > \mathcal{O}(10^{-2})$ exceeds unity. 
On the other hand, absorption signals smoothed at lower resolution show a reduced abundance at larger optical depth.

The middle and right panels of Figure \ref{fig:dNdlntdz:MHFidz10:res} shows how the total abundance depends on halo masses with the resolution $\Delta \nu \approx 5.4\mr{kHz}$ and $54\mr{Hz}$, respectively. 
The top panels depict $\mrd^2 N/\mrd \ln \tau \mrd z$, while the bottom panels show the ratio of each halo abundance to the total. 
Each color in the figures represents a specific range of the halo mass, in the unit of $\Msunhinv$. For example, $\mathcal{O}(10^6)$ indicates a mass range of $10^6 \Msunhinv \leq M_\mr{vir} < 10^7 \Msunhinv$. 

Both results of $\Delta \nu \approx 5.4 \mr{kHz}$ and $54 \mr{Hz}$ indicate that the smaller minihalos significantly contribute to the larger optical depth and the greater abundance. 
Almost all the optical depths with $\tau > 10^{-2}$ reflect minihalos with masses on the order of $M_\vir = \mathcal{O}(10^5)\Msunhinv$. 
Notably, minihalos with masses around $\mathcal{O}(10^5)\Msunhinv$ and $\mathcal{O}(10^6)\Msunhinv$ contribute roughly equally to optical depths near $\tau = 10^{-2}$. 

It is important to note that only minihalos with masses smaller than $\mathcal{O}(10^{6})\Msunhinv$ can produce $\tau > \mathcal{O}(10^{-2})$. 
Optical depths with $\tau \lesssim 10^{-3}$ originate from both the higher mass-minihalos and the wings of the profile, and are not significant, as the optical depth contribution from the IGM increases in that region (see Subsection \ref{subsec:withigm}). 
Furthermore, we observe that the range of the optical depths for minihalos with masses of $\mathcal{O}(10^5)\Msunhinv$ and $\mathcal{O}(10^6)\Msunhinv$ broadens in the region of $\tau > \mathcal{O}(10^{-2})$ for $\Delta \nu \approx 54 \mr{Hz}$ compared to $\Delta \nu \approx 5.4 \mr{kHz}$. 
This trend is also attributed to the finer resolution, which allows for a better resolution of the interiors of the halos, resulting in larger optical depths, similar to what is observed in the left panel. 
In summary, the presence the absorption lines with $\tau > \mathcal{O}(10^{-2})$ indicates the existance of minihalos with masses of $\mathcal{O}(10^5) \Msunhinv$ and $\mathcal{O}(10^6) \Msunhinv$. These numbers exceed unity when we consider a frequency resolution of $\Delta \nu \lesssim 100 \mr{Hz}$, which is sufficient to resolve these minihalos.

\subsection{Comparison with ``diffuse" IGM}
\label{subsec:withigm}

In this subsection, we compare the contributions to optical depth from minihalos and the diffuse IGM. 

\begin{figure}[htbp]
\centering
\includegraphics[width=\textwidth]{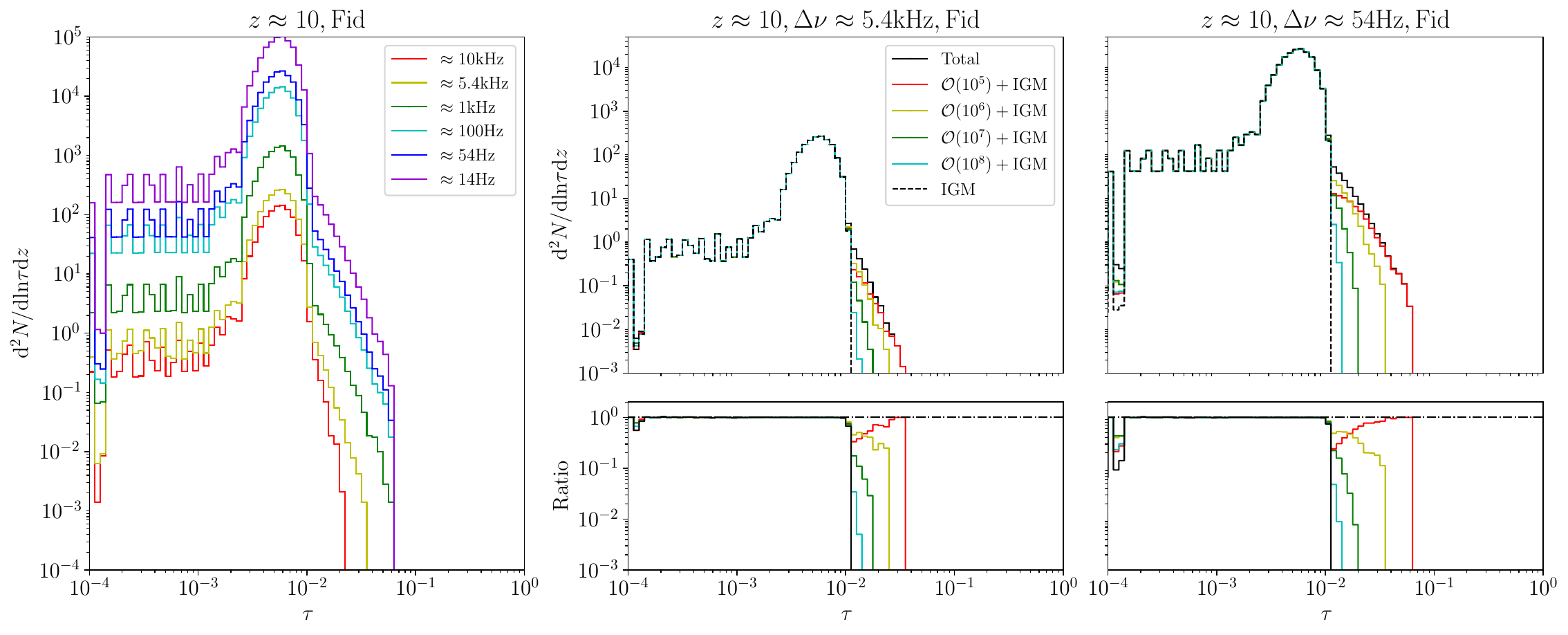}
\caption{
Same as Figure \ref{fig:dNdlntdz:MHFidz10:res}, but we include the contribution from the "diffuse" IGM to the optical depth including the results for each halo mass bin.
In the middle and right panels, we separately depict the contribution from the diffuse IGM with the dashed line, which largely overlaps with the other colored lines.
}
\label{fig:dNdlntdz:MHexIGMFidz10:res}
\end{figure}

Figure \ref{fig:dNdlntdz:MHexIGMFidz10:res} shows the abundance of absorption lines computed with both the contributions from minihalos and the diffuse IGM. 
The left panel presents results with different $\Delta \nu$, and the middle and right panels show the dependence of the total abundance on different halo mass ranges and the diffuse IGM with $\Delta \nu \approx 5.4 \mr{kHz}$ and $54 \mr{Hz}$, respectively. 
The diffuse IGM contribution is described by the dashed lines. 

For optical depths $\tau \lesssim 10^{-2}$, every line nearly overlaps, indicating that the contribution in that region primarily comes from the diffuse IGM. 
We can roughly confirm this scale using the average value from the simulation shown in the middle panel of Figure \ref{fig:xHI_T_Frac_history:fid} and equation (\ref{tauofIGM:eq}). 
On the other hand, the halo contribution becomes dominant in the region where $\tau > \mathcal{O}(10^{-2})$, which is clearly shown in the middle and right panels in Figure \ref{fig:dNdlntdz:MHexIGMFidz10:res}. 
In our fiducial analysis of the 21cm forest spectrum at redshift $z\approx 10$,  we may conclude that absorption lines with optical depths $\tau \lesssim 10^{-2}$ indicate the presence of diffuse IGM, while $\tau \gtrsim 10^{-2}$ signifies the existence of minihalos.  Almost all of these minihalos have masses on the order of $\mathcal{O}(10^5) \Msunhinv$ with some being approximately $\mathcal{O}(10^6) \Msunhinv$. 
It is important to note that this result is contingent under the conditions of reionization and the thermal history illustrated in Figure \ref{fig:xHI_T_Frac_history:fid} (see also a discussion in Subsection \ref{subsec: resultwithIGM}). 

As explained in relation to Figure \ref{fig:dNdlntdz:MHFidz10:res}, the abundance is larger with finer frequency resolution. 
When the diffuse IGM contribution is included, the optical depth associated with minihalo absorption is slightly enhanced by the overlapping absorption from the diffuse IGM. This overlap occurs because the peculiar velocities of minihalos Doppler-shift their absorption profiles into frequency ranges where diffuse IGM absorption is also present, as illustrated in Figure \ref{fig:spectrumex2}. Consequently, the optical-depth distribution shifts slightly toward higher values, leading to an apparent increase in the abundance at the high-optical-depth end.
Nevertheless, 
the number of absorption lines with $\tau > \mathcal{O}(10^{-2})$ exceeds unity with the frequency resolution $\Delta \nu \lesssim 100 \mr{Hz}$, 
whereas the optical depth by the diffuse IGM is abundant even with lower resolutions. 
\subsection{Redshift dependence}

As illustrated in Figure \ref{fig:xHI_T_Frac_history:fid}, the fraction of the HI in the gas surrounding minihalos, as well as the gas temperature, changes with redshift. 
In this subsection, we explore the implications of this redshift-dependent environment. 
Here we examine five distinct phases: 
\begin{itemize}
\item[(i)] before the epoch of heating (EoH), when X-ray gas heating to the IGM becomes effective (approximately at redshift $z \approx 20$);
\item[(\romaii)] the onset of EoH (approximately at redshift $z \approx 15$);
\item[(\romaiii)] around the beginning of the EoR, with neutral hydrogen fraction $x_\mr{HI} \approx 0.961$ (approximately at redshift $z \approx 10$);
\item[(\romaiv)] the middle stage of the EoR when the neutral hydrogen fraction drops to $x_\mr{HI} \approx 0.890$ and the Gaseous halo ratio reaches about $\sim 70 \%$ (approximately at redshift $z \approx 8.5$);
\item[(v)] the later middle stage of the EoR, when the neutral hydrogen fraction is $x_\mr{HI} \approx 0.706$, and the ratio of Gaseous halos and Dead halos are both around $\sim 50 \%$ (approximately at redshift $z \approx 7$). 
\end{itemize}

\begin{figure}[htbp]
\centering
\includegraphics[width=1.0\textwidth]{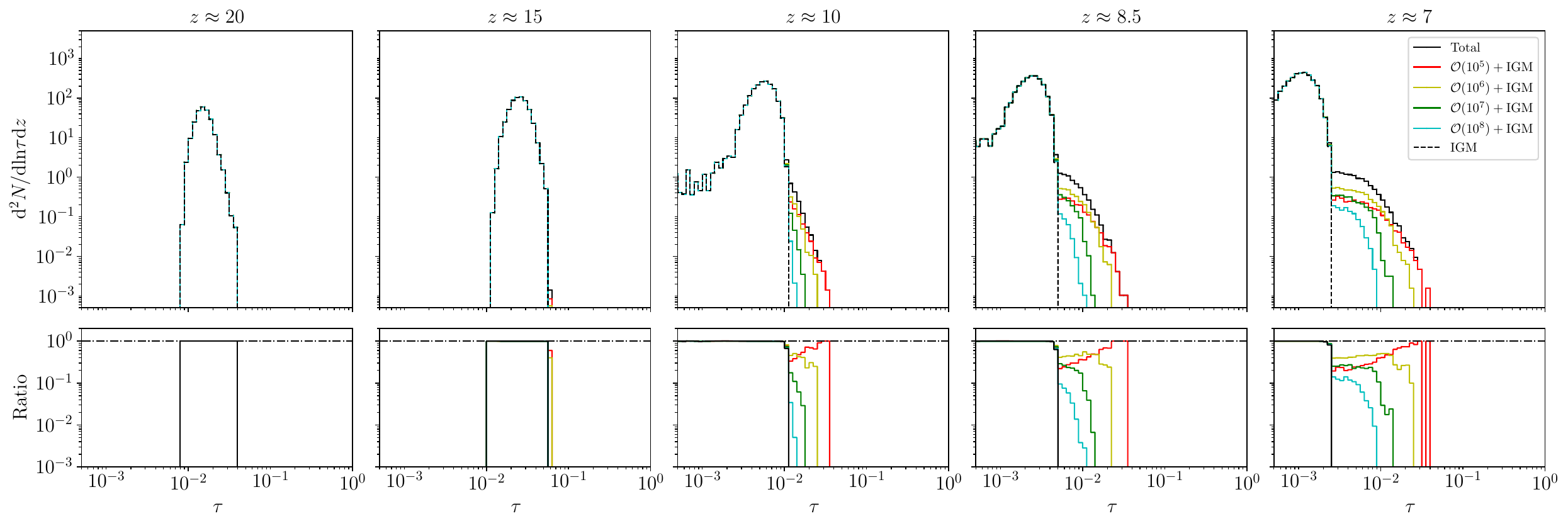}
\caption{
Redshift evolution of the optical-depth distribution from $z \approx 20$ (stage (i)) to $z \approx 7$ (stage (v)), showing the distributions obtained by combining the diffuse IGM contribution with minihalos in different mass ranges. Frequency resolution of the analysis is $\Delta \nu \approx 5.4 \mr{kHz}$. 
The color representation and the unit of the halo mass ranges are the same as those used in the middle and right panels of Figure \ref{fig:dNdlntdz:MHexIGMFidz10:res}.
}
\label{fig:dNdlntdz:MHexIGMFidzall5.4khz:MhRatio}
\end{figure}

\begin{figure}[htbp]
\centering
\includegraphics[width=1.0\textwidth]{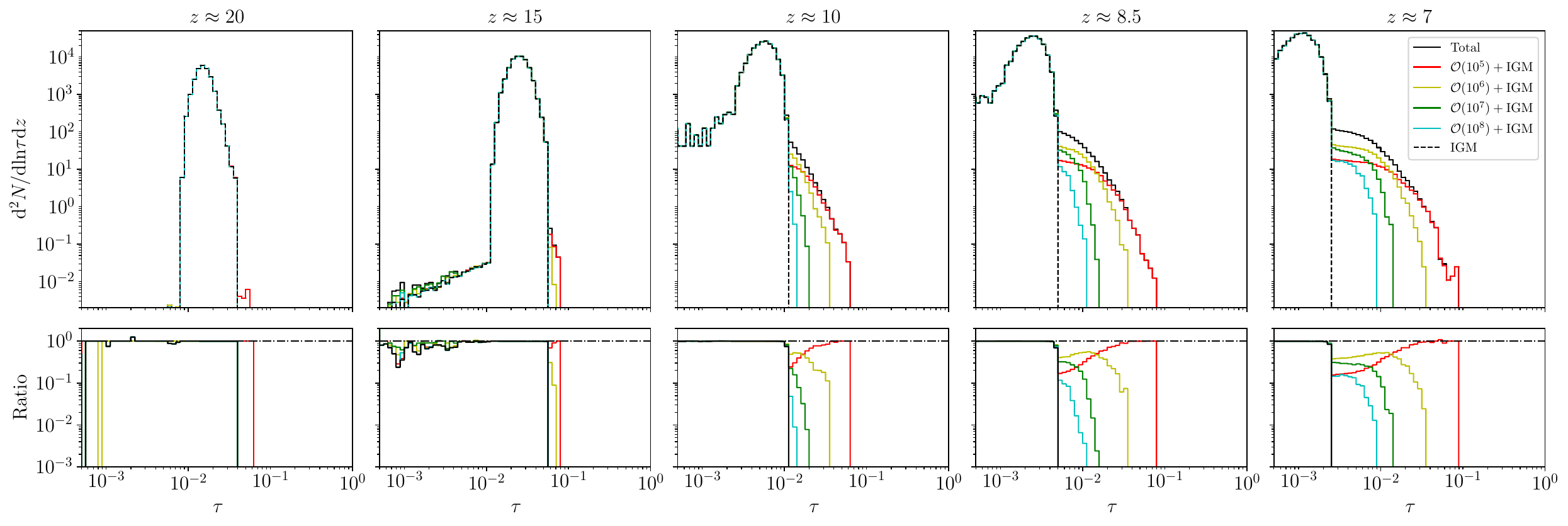}
\caption{
Same as Figure \ref{fig:dNdlntdz:MHexIGMFidzall5.4khz:MhRatio}, but for $\Delta \nu \approx 54 \mr{Hz}$. 
}
\label{fig:dNdlntdz:MHexIGMFidzall54hz:MhRatio}
\end{figure}

Figure \ref{fig:dNdlntdz:MHexIGMFidzall5.4khz:MhRatio} and \ref{fig:dNdlntdz:MHexIGMFidzall54hz:MhRatio} show that the redshift evolution of the abundance of the optical depth and its contribution from each halo mass range,
with $\Delta \nu \approx 5.4 \mr{kHz}$ and $\Delta \nu \approx 54 \mr{Hz}$, respectively. 
From left to right, each panel represents results from (i) to (v). 
The line representations are consistent with those in the middle and right panels of Figure \ref{fig:dNdlntdz:MHexIGMFidz10:res}. 
At high redshift before EoH such as (i) and (ii), the temperature of the diffuse IGM is low, resulting in an increase in the optical depth of the diffuse IGM, reaching up to a few $10^{-2}$. 
This leads to the conclusion that at these redshifts
the IGM optical depth dominates even in the region of $\tau > \mathcal{O}(10^{-2})$. Additionally, only very small fraction of contributions from $\mathcal{O}(10^5) \Msunhinv$ minihalos is distinctive. 
Number of absorption lines by minihalos is only $\mathcal{O}(10^{-3})$ for $\Delta \nu \approx 5.4 \mr{kHz}$ and $\mathcal{O}(10^{-2})$ for $\Delta \nu \approx 54 \mr{Hz}$ at $z \approx 15$.
There is almost no appearance with $\Delta \nu \approx 5.4 \mr{kHz}$, and $\mathcal{O}(10^{-3})$ with $\Delta \nu \approx 54 \mr{Hz}$ at $z \approx 20$.

In contrast, at lower redshifts when X-ray heating to the diffuse IGM is effective in the middle of EoH, and when sufficient gaseous halos still exist during the EoR, the optical depth caused by minihalos becomes quite apparent. 
As the IGM temperature increases due to heating and the reionization proceeds (indicated by a decrease in $x_\mr{HI}$), the optical depth by the diffuse IGM decreases. However, the optical depth contributed by minihalos remains significant, even in regions where $\tau < 10^{-2}$ to some extent. 
Here, we can see the subtle excess of 1 around $\tau \sim 5 \times 10^{-2}$ in the Ratio $z \approx 7$ in Figure \ref{fig:dNdlntdz:MHexIGMFidzall54hz:MhRatio}. 
This is because there are the pixel where the optical depths by multihalos are summed, like Figure \ref{fig:schematicview} shows. 
The total optical depth in "Total" is divided by the optical depth by each mass halos, and the number of smaller optical depth by each halo gets larger than total like $\tau \sim 5-6 \times 10^{-2}$ in the Figure. 

Eventually, we observe that an optical depth greater than $\tau > \mathcal{O}(10^{-2})$ indicates the influence of the minihalos, which primarily have masses around $\mathcal{O}(10^5) \Msunhinv$ and, to a lesser extent, around $\mathcal{O}(10^6) \Msunhinv$. This situation occurs when the optical depth contribution from the diffuse IGM  decreases due to heating and ionization. Conversely, when X-ray heating is ineffective and the diffuse IGM temperature is low, the optical depth from minihalos becomes indistinguishable from that from the diffuse IGM. 

\subsection{
Impact of the IGM thermal history}
\label{subsec: resultwithIGM}

The value of the astronomical parameter, particularly regarding the amount of X-rays, has been frequently discussed in previous studies, including those related to the 21 cm forest (e.g. \cite{2011MNRAS.410.2025X,2012MNRAS.425.2988M,2013MNRAS.428.1755C,2023NatAs...7.1116S}) 
.  This parameter is uncertain in the EoR,  yet it plays an important role in understanding the thermal and reionization history of the universe. 
In this study, we also investigate the effect by the parameter, $L_{X<2\mr{keV}}\;[\mr{erg\;s^{-1}\;\Msun^{-1}\;\mr{yr}}]$, the total integrated soft X-ray luminosity per SFR escaping into the IGM, which normalizes the specific X-ray luminosity per SFR \cite{2020MNRAS.495..123Q,2023MNRAS.520.3368B,2025MNRAS.540..483V}. 
We set our fiducial value as $L_\mr{X<2keV}/\mr{SFR} = 3.16 \times 10^{40}$. 
Additionally, we defined the HighX model as $L_\mr{X<2keV}/\mr{SFR} = 3.16 \times 10^{41}$, and the LowX model as $L_\mr{X<2keV}/\mr{SFR} = 3.16 \times 10^{39}$. 

\begin{figure}[htbp]
\centering
\includegraphics[width=1.0\textwidth]{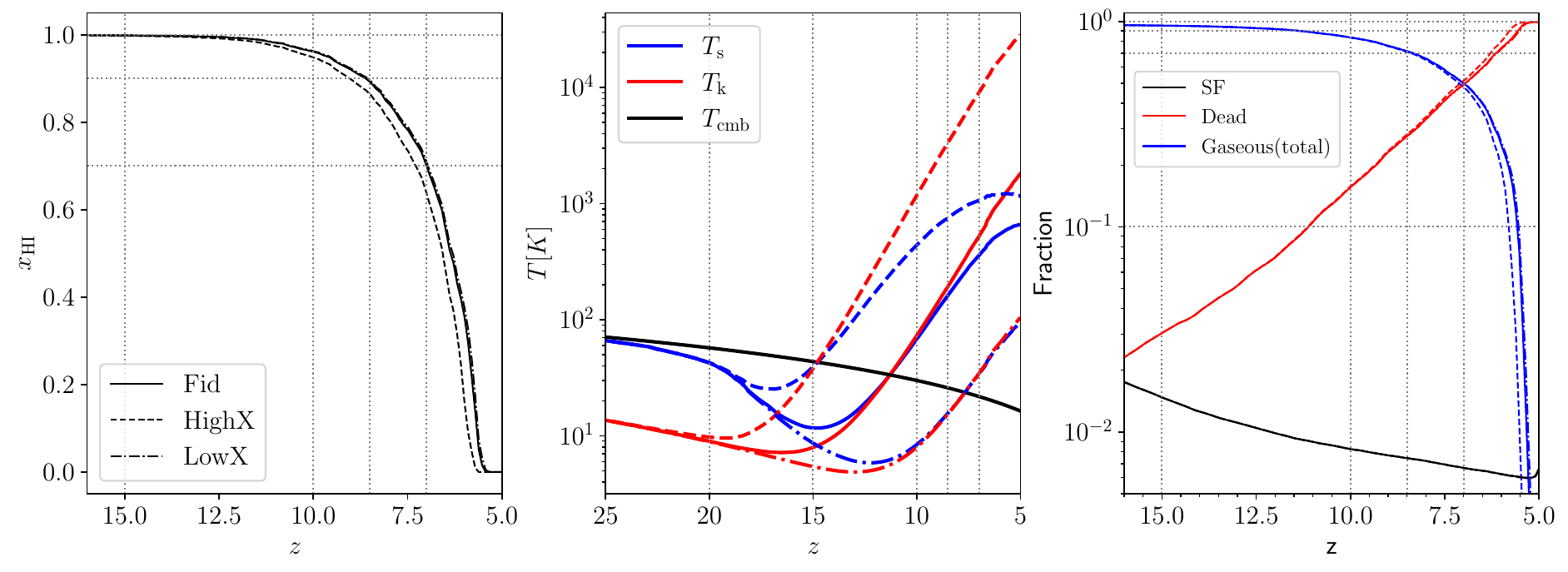}
\caption{
The histories of the HI fraction $x_\mr{HI}$ (left), the volume-averaged temperatures (middle), and the halo classification (right). 
The solid lines represent the fiducial model with $L_\mr{X<2keV}/\mr{SFR}\; [\mr{erg\;s^{-1}\;\Msun^{-1}\;\mr{yr}}] = 3.16 \times 10^{40}$. The dashed lines correspond to the HighX model with $L_\mr{X<2keV}/\mr{SFR} = 3.16 \times 10^{41}$, and the dot-dashed lines represent the LowX model with $L_\mr{X<2keV}/\mr{SFR} = 3.16 \times 10^{39}$. 
}
\label{fig:xHI_T_Frac_history:hfl}
\end{figure}

Figure \ref{fig:xHI_T_Frac_history:hfl} shows the HI fraction history, the thermal history, and the fractions of three types of halos in the simulation for the three models, from left to right. 
The resulting line for the fiducial model is identical to that shown in Figure \ref{fig:xHI_T_Frac_history:fid}. 
From the left panel, it can be observed that HighX shows relatively early ionization compared to the others. However, all of their end of the EoR occur at $z < 6$, indicating late reionization.

In the middle panel, there are notable differences among the three models. The 
HighX model results in the early heating of the IGM before the Wouthuysen-Field coupling \cite{1952AJ.....57R..31W,1958PIRE...46..240F} is fully established. In contrast, the LowX model resembles so-called the "cold" reionization scenario, in which EoH is delayed due to insufficient X-ray heating at high redshift, resulting in the EoR overlaping with the EoH \cite{2013MNRAS.431..621M, 2017MNRAS.472.2651G}. 

The right panel shows that the numbers of the three types of the halos rarely change within the three models until reionization progresses to some extent, i.e., $z \sim 7$. 
After this point, the number of gaseous halos slightly decreases while that of dead halos increases in the HighX model.
As mentioned regarding the parameter definition, $L_{X<2\mr{keV}}$ controls the amount of X-ray radiation affecting the IGM by escaping from halos. 
This parameter does not alter the total number of the halos; however, the numbers of gaseous and dead halos exhibit slight changes according to the differences in reionization history. This is evident from equation (\ref{equation:mfilt}), where the mass criterion $M_\mr{filt}$ depends on the reionization redshift $z_\mr{ion}$. 

\begin{figure}[htbp]
\centering
\includegraphics[width=\textwidth]{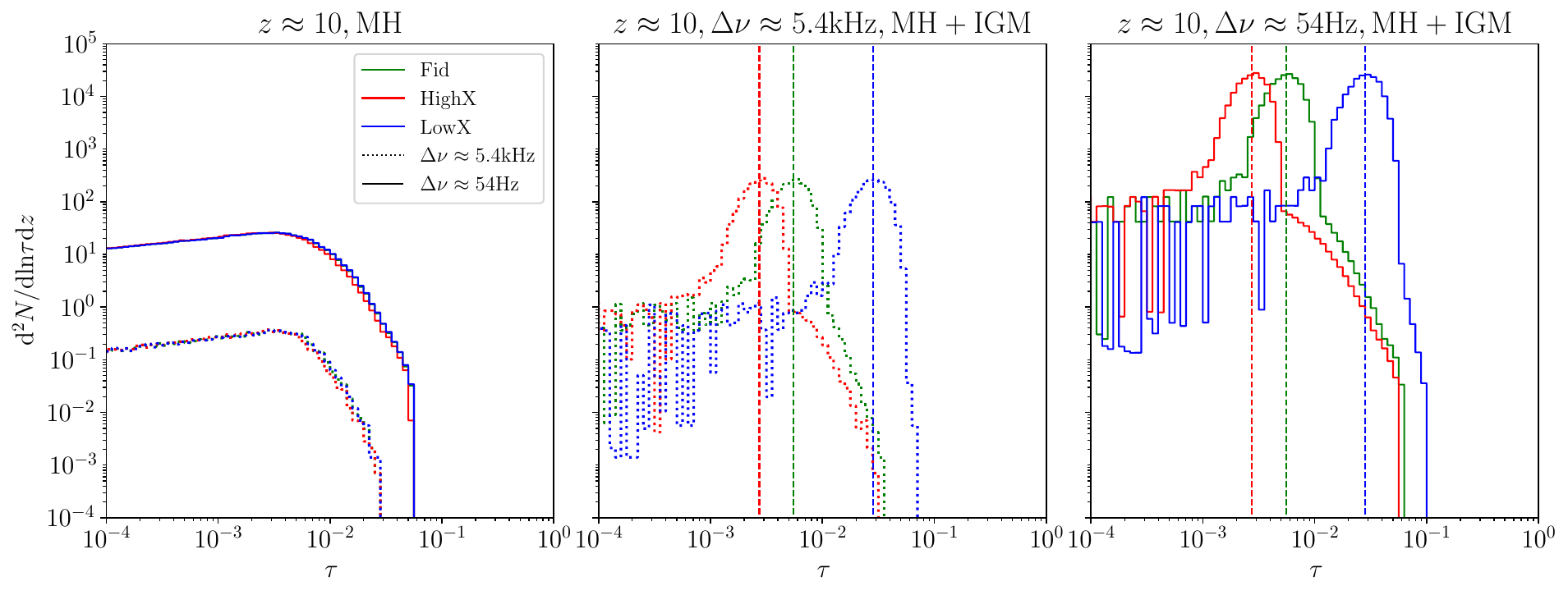}
\caption{
Left: $\mrd^2 N/ \mrd \ln \tau \mrd z$ at $z \approx 10.0$ of the optical depth by halos for the $L_\mr{X<2keV}/\mr{SFR}$ parameter with $\Delta \nu \approx 5.4\mr{kHz}$ (dot) and $54\mr{Hz}$ (solid). 
The green lines represent the fiducial model, the red lines represent the HighX model, and the blue lines represent the LowX model. 
In each frequency resolution, the three kinds of lines are roughly overlapping
. 
Middle: $\mrd N^2/ \mrd \ln \tau \mrd z$ at $z \approx 10.0$ of the optical depth computed with both the minihalos and the diffuse IGM for the parameter $L_\mr{X<2keV}/\mr{SFR}$ with $\Delta \nu \approx 5.4\mr{kHz}$.
As in the left panel, different colors represent different X-ray heating models as indicated. 
The vertical dashed lines represent the average of all optical depths. 
Right: the same as the middle panel but with $\Delta \nu \approx 54\mr{Hz}$. 
}
\label{fig:dNdlntdz:MHFHLz10}
\end{figure}

Figure \ref{fig:dNdlntdz:MHFHLz10} illustrates how these differences in thermal and reionization histories affect the $21$ cm optical depth distribution at $z\approx 10$.
The left panel shows the contribution from minihalos alone at spectral resolution of  $\Delta\nu \approx 5.4~\mathrm{kHz}$ and $54~\mathrm{Hz}$. As expected from the similar halo populations in the three models, the optical-depth distributions are nearly identical.
In contrast, the middle and the right panels, which include the diffuse IGM contribution, show clear 
dependence on the X-ray luminosity.
A higher $L_\mr{X<2keV}/\mr{SFR}$ leads to earlier IGM heating and hence the lower optical depth of the diffuse IGM. Conversely, the colder IGM produces stronger absorption, extending the optical-depth distribution toward the region dominated by minihalos in the fiducial model.

The vertical dashed lines in Figure \ref{fig:dNdlntdz:MHFHLz10} 
represent the mean optical depths of the combined minihalos and diffuse IGM spectra. These values are almost identical to the mean optical depths of the IGM alone, indicating that minihalos contribute little to the mean absorption.
For example, in the fiducial model, the mean optical depth with and without minihalos are $5.534 \times 10^{-3}$ and $5.521 \times 10^{-3}$, respectively.
As discussed by Mack \& Wyithe (2012) \cite{2012MNRAS.425.2988M}, the mean optical depth is sensitive to the X-ray heating history. Our results further demonstrate that this heating history determines the optical-depth range over which minihalo absorption can be distinguished from that of the diffuse IGM.
We found that if the IGM temperature is low enough, as in the LowX model, absorption lines from minihalos are indistinguishable from those created by the diffuse IGM in the range $\tau \gtrsim \mathcal{O}(10^{-2})$. 

Therefore, to effectively detect minihalos directly, we need to understand the thermal history of the IGM, which is key to separate the $\tau$ contributions from minihalos and the IGM. 
Because we focus on the period around the beginning of EoR when the gas is still almost neutral, the reionization history is not critical for this detectability. 

\subsection{The influence of halo thermal profile}
In the subsections above, we investigate the contribution of minihalos to the 21 cm optical depth, assuming that the temperature distribution of halos is isothermal at the virial temperature. 
However, the temperature profile of a halo is not necessarily isothermal. 
When a halo collapses adiabatically, the central core region is expected to become hotter, while the outer regions remain cooler.
In this subsection, we consider an alternative halo temperature profile: we keep to assume the halo DM and gas 
density profiles the same as in the above sections, but the 
temperature profiles are determined adiabatically, assuming hydrostatic equilibrium with its gravity, or the adiabatic profile.  

\begin{figure}[htbp]
\centering
\includegraphics[width=0.8\textwidth]{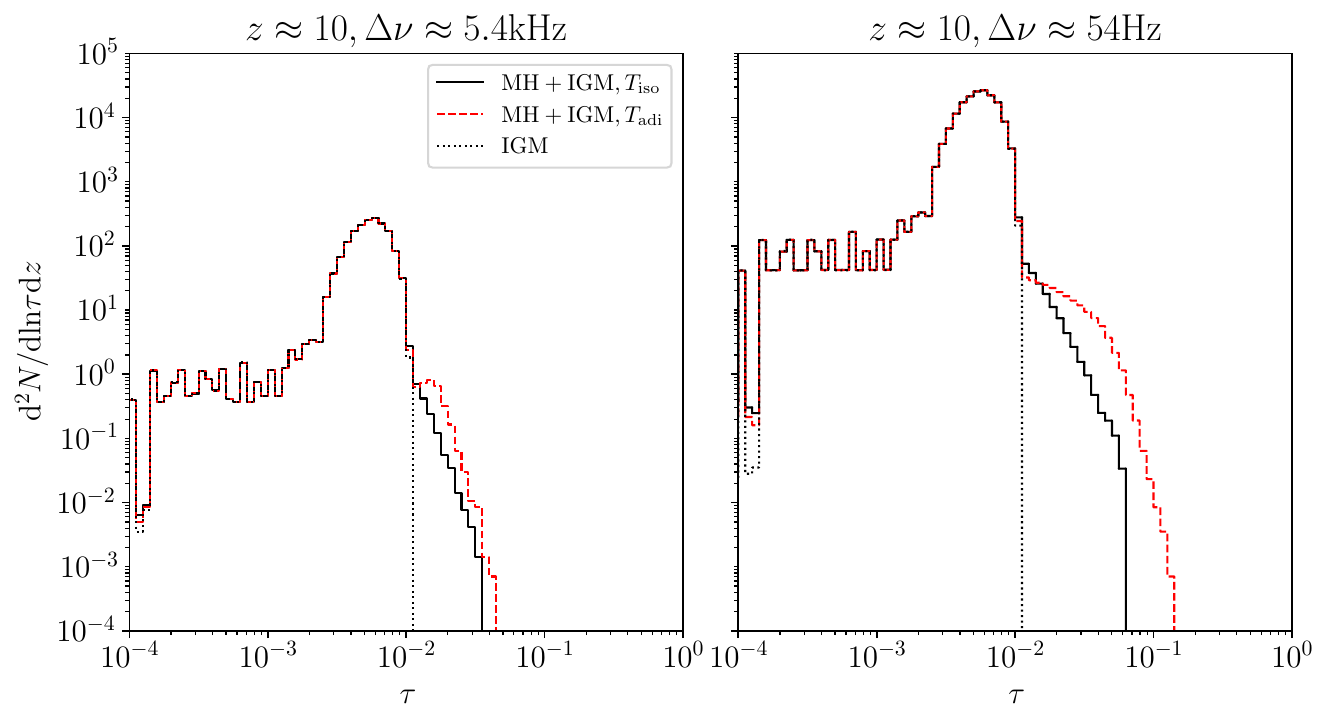}
\caption{
$\mrd N^2/ \mrd \ln \tau \mrd z$ at $z \approx 10.0$ of the optical depth by minihalos and the diffuse IGM in the fiducial model for the halo temperature profiles, with $\Delta \nu \approx 5.4 \mr{kHz}$ (left) and $54\mr{Hz}$ (right). 
The black solid line shows the result with the isothermal profile at the virial temperature, and the red dashed line shows the result with the adiabatic temperature profile. 
The dotted line shows the optical depth by the diffuse IGM. 
}
\label{fig:dNdlntdz:MHexIGMFidz10:Tadi}
\end{figure}

The equation of hydrostatic equilibrium is written as 

\begin{align}
    \label{HSE:eq}
    &\frac{1}{\rho_\mr{g}(r)} \frac{\mrd }{\mrd r} \left( T_\mr{g}(r) \rho_\mr{g}(r) \right) = - \kappa \frac{M_\mr{tot}(r)}{r^2} \notag \\ 
    \Leftrightarrow&\rho_\mr{g}(r) \frac{\mrd T_\mr{g}(r)}{\mrd r} + T_\mr{g}(r) \frac{\mrd \rho_\mr{g}(r)}{\mrd r} = - \kappa \frac{M_\mr{tot}(r)}{r^2}\rho_\mr{g}(r), 
\end{align}
where $\kappa = \mu m_\mr{p} G/k$, and 
$\rho_\mr{g}(r)$, $T_\mr{g}(r)$, and $M_\mr{tot}(r)$ are the gas density, gas temperature, and total mass (including DM and gas) at the radius of the halo $r$, respectively. 
Therefore, when comparing the isothermal and adiabatic profiles under the same gravity, as shown in the LHS of equation (\ref{HSE:eq}), the isothermal temperature supports only the second term, whereas the adiabatic temperature includes the temperature-gradient term. 
This means that the isothermal $T_\vir$ exceeds the adiabatic temperature.
In other words, the adiabatic profile model has a lower temperature, resulting in a larger optical depth and greater detectability of minihalos, as shown in Figure \ref{fig:dNdlntdz:MHexIGMFidz10:Tadi}. 

Figure \ref{fig:dNdlntdz:MHexIGMFidz10:Tadi} shows the abundance of optical depth from minihalos and the diffuse IGM for different halo temperature profiles. 
The black solid line denotes the result from the isothermal temperature profile, which matches the fiducial model in the middle and right panels of Figure \ref{fig:dNdlntdz:MHFHLz10}.
The red dashed line shows the result from the adiabatic profile, and the dotted line shows that from the diffuse IGM. 
As for the minihalo contribution, the optical depth from the adiabatic profile is generally larger than that from the isothermal profile.  Both its maximum value and the number of contributing to high optical depth $\tau \gtrsim 10^{-2}$ are greater for the adiabatic profile case.

\section{Discussion}
\label{sec: discussion}
In the sections above, we investigate the contribution of minihalos to the 21 cm forest and compare it with that of the diffuse IGM. 
It should be noted that we neglected the contribution of the moderate overdensity structures with $\Delta>10$, namely filaments. 
Because filaments are non-virialized structures, a hydrodynamic simulation is needed, especially for the internal temperature distribution, to estimate their optical depth. 
However, according to Carilli et al. (2002) \cite{2002ApJ...577...22C}, the filaments have approximately $\delta \sim 10$, and the spin temperature exceeds $100 \mr{K}$, even though the volume-averaged 
temperature history differs from ours. 
Using the approximate scaling $\tau \propto \delta/ T$ around the beginning of the EoR ($x_\mr{HI} \sim 1$), we can compare the optical depths of filaments, the diffuse IGM and minihalos. In our fiducial model, the diffuse IGM has $\delta = \mathcal{O}(1)$ and $T_\mr{S} \sim 70 \mr{K}$ while minihalos with $M_\mr{vir} = 10^5 \Msunhinv$ have $\delta = \mathcal{O}(100)$ and $T_\mr{vir} \approx 300 \mr{K}$ at $z \approx 10$. This estimate suggests that the optical depth of filaments would be a few times larger than that of the diffuse IGM but lower than that of minihalos. 
Although hydrodynamic simulations are needed for a precise analysis, this comparison suggests that the contribution to optical depths of $\tau > \mathcal{O}(10^{-2})$ is dominated by minihalos. 

Additionally, the $H_2$ self-shielding effect (e.g. \cite{2001ApJ...560..580R, 2009MNRAS.395.1280H,2015MNRAS.452.1233H,2021MNRAS.507.1775S}), which we do not consider in this study, may affect our results. 
Self-shielding suppresses $H_2$ dissociation inside halos and promotes star formation through $H_2$ molecular cooling. 
An increase in the number of SF halos can enhance the intensity of the background radiation field and change the thermal history, potentially altering the minihalo contribution to the optical depth in the 21 cm forest. 
Moreover, even before star formation begins, the gas profile may be modified when the $H_2$ shielding is effective \cite{2009MNRAS.395.1280H}. 
We leave these influences of the self-shielding effect on the minihalo absorption lines in the 21 cm forest for future work. 

As for the prospect of future feasibility of detection, we follow \cite{2002ApJ...579....1F,2015MNRAS.453..101C,2023JCAP...03..017K,2021MNRAS.506.5818S,2023MNRAS.519.3027S} and express the minimum required flux density $S_\mr{min}$ as
\begin{equation}
    \label{smin:eq}
    S_\mr{min} = 10.3 \mr{mJy}\left( \frac{0.01}{\tau} \right) \left( \frac{S/N}{5}\right)\left( \frac{5\mr{kHz}}{\Delta \nu}\right)^{1/2}\left( \frac{1000 \mr{hr}}{t_\mr{int}}\right)^{1/2}\left( \frac{1000 \mr{m^2 K^{-1}}}{A_\mr{eff}/T_\mr{sys}}\right), 
\end{equation}
where $S/N$ is a signal-to-noise ratio, $t_\mr{int}$ is the integration time, $T_\mr{sys}$ is the system temperature, and $A_\mr{eff}$ is the effective area of the telescope. 
If we assume the SKA2 sensitivity $A_\mr{eff}/T_\mr{sys} = 5000 \mr{m^2 K^{-1}}$ \cite{2019arXiv191212699B} and $S/N=5$ and $t_\mr{int} = 100 \mr{hr}$, the background source flux density $S_\mr{min} \approx 46.1 \mr{mJy}$ is needed for the detection of the optical depth $\tau = 10^{-2}$ with $\Delta \nu = 100 \mr{Hz}$. 
Conversely, for a background flux density of $S_\mr{min} = 10.3 \mr{mJy}$ and the same values of $A_\mr{eff}$ and $S/N$ and $t_\mr{int}$, the minimum optical depth to be detected is $\tau \approx 4.47 \times 10^{-2}$ with $\Delta \nu = 100 \mr{Hz}$. Figure \ref{fig:dNdlntdz:MHexIGMFidz10:res} shows that the abundance at this optical depth with $\Delta \nu \approx 100 \mr{Hz}$ is below unity. Therefore, detecting minihalos at this threshold would require a brighter background source.

\begin{figure}[htbp]
\centering
\includegraphics[scale=0.6]{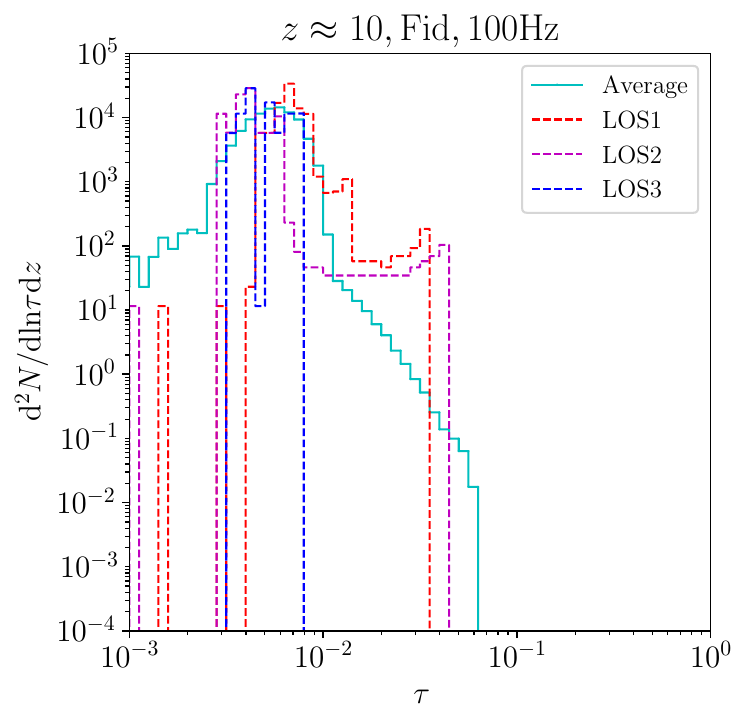}
\hfill
\includegraphics[scale=0.6]{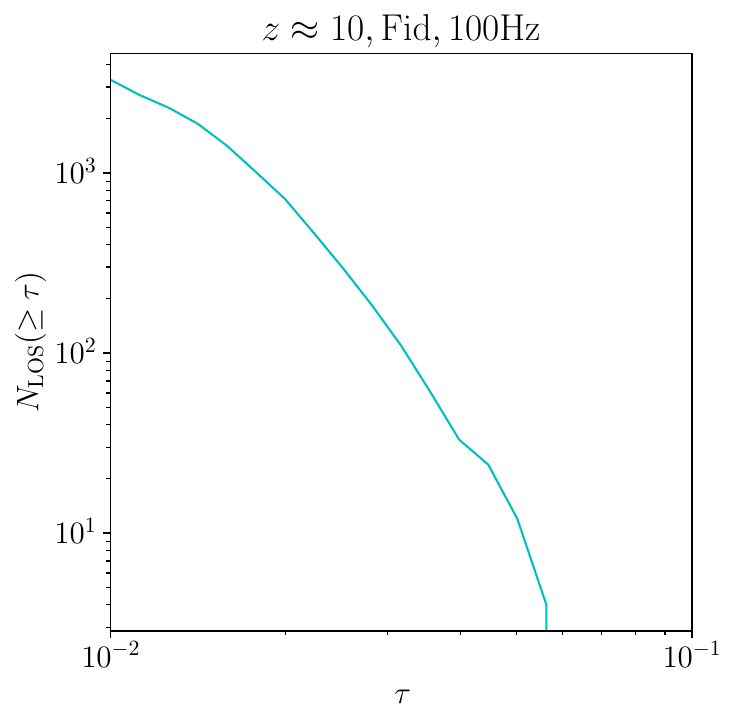}
\caption{
Left: The abundance of the optical depth $\mrd N^2/ \mrd \ln \tau \mrd z$ at $z \approx 10$ and the fiducial model with the frequency resolution $\Delta \nu \approx 100 \mr{Hz}$.
"Average" is the same result as $100 \mr{Hz}$ in the left panel of Figure \ref{fig:dNdlntdz:MHexIGMFidz10:res}. 
"LOS1" to "LOS3" show the results for individual LOS as examples. 
LOS1 crosses four halos, LOS2 crosses one halo, and LOS3 crosses no halos. 
Right: The cumulative number of LOSs $N_\mr{LOS}(\geq \tau)$ which have the optical depth greater than or equal to the threshold $\tau$, at $z \approx 10$ and the fiducial model with the frequency resolution $\Delta \nu \approx 100 \mr{Hz}$.
The result is the number out of a maximum of $128^2$ LOSs we take in this study. 
}
\label{fig:indiv}
\end{figure}

However, the optical-depth distribution presented in Section \ref{sec:results} is the result of taking the statistical average of all LOSs, whereas actual observations probe individual LoSs toward background radio sources.
An advantage of our simulation-based approach is that it allows us to investigate the LOS-to-LOS variations in minihalo absorption.
The left panel of Figure \ref{fig:indiv} illustrates this validation by comparing the averaged optical depth distribution with those along three individual LoS that intersect different numbers of minihalos. Although strong absorption features are rare on average, individual LoSs can exhibit substantially enhanced abundances at high optical depths.
We note that, at a spectral resolution of $\Delta\nu \approx 100~\mathrm{Hz}$, a single minihalo can contribute to multiple frequency channels. Nevertheless, these results suggest that strong minihalo absorption features could be detectable along favorable LOSs toward background radio sources with flux densities as low as $10.3~\mathrm{mJy}$ under the observational conditions considered above. This highlights the importance of accounting for LOS-to-LOS variations when assessing the detectability of the 21 cm forest.

Finally, the right panel of Figure \ref{fig:indiv} shows the cumulative number of LOSs where the pixel have the optical depth larger than or equal to the threshold $\tau$, at $z \approx 10$ with $\Delta \nu \approx 100 \mr{Hz}$.  
For example, the number of LOSs which have $\tau \geq 4.47 \times 10^{-2}$ is 24, those who have $\tau \geq 3.0 \times 10^{-2}$ is 139, and those who have $\tau \geq 2.0 \times 10^{-2}$ is 713, out of a maximum of $128^2$. 
Their fractions to total are $1.465\times10^{-3}$, $8.484\times10^{-3}$, and $4.352\times10^{-2}$, respectively. 
Our simulation boxsize is $L_\mr{box} = 10 \mr{cMpc}/h$ on a side, so that the estimated field of view (FoV) by $\Delta\theta^2_\mr{FoV}(z) = (L_\mr{box}/D_\mr{c}(z))^2$ is $\sim 5.27^2 \;\mr{arcmin}^2$ at $z = 10$, where $D_\mr{c}(z)$ is the comoving distance. 
This result plays a role as a reference for estimating the feasibility of detection. 

\section{Summary}
\label{sec: summary}
 
In this paper, we model how minihalos contribute to the optical depth in the 21 cm forest around the start of the EoR. 
To accurately estimate the contribution of minihalos, we use the spatial and mass distributions of halos from the semi-analytical simulation \textit{MERAXES}, which resolves halos down to masses of $\sim 2 \times 10^5 \Msunhinv$ taking the background radiation feedback into consideration. 
After pasting an analytical halo gas density profile and a isothermal temperature of $T_\mr{vir}$ to each halo at its simulated location, we calculate the optical depth spectra along $128^2$ LOSs at frequency resolutions sufficient to resolve the internal structure of minihalos. 

We estimate the optical depth abundance $\mrd^2 N/\mrd \ln{\tau} \mrd z$ and find that optical depths $\tau > \mathcal{O}(10^{-2})$ arise primarily from minihalos with masses of $\mathcal{O}(10^5) \Msunhinv$ and partially $\mathcal{O}(10^6) \Msunhinv$ around $\tau \sim 10^{-2}$. 
With the frequency resolution as high as $\Delta \nu \lesssim 100 \mr{Hz}$, the abundance of these absorption features exceeds unity at the onset of the EoR. 
On the other hand, weaker absorption features with $\tau < \mathcal{O}(10^{-2})$ arise primarily from the ``diffuse" IGM, defined here as gas with an overdensity of $\Delta \lesssim 10$, and are more abundant than those produced by the minihalos. 

However, this separation above depends strongly on the IGM temperature history. 
When the IGM temperature is low enough to have optical depth as high as $\tau \gtrsim 10^{-2}$, as occurs before the EoH in the fiducial model or in the LowX model, absorption by minihalos is indistinguishable from that by the diffuse IGM.

\acknowledgments
The simulation of this work was performed on the OzSTAR national facility at Swinburne University of Technology. The OzSTAR program receives funding in part from the Astronomy National Collaborative Research Infrastructure Strategy (NCRIS) allocation provided by the Australian Government, and from the Victorian Higher Education State Investment Fund (VHESIF) provided by the Victorian Government. 
The analysis of this work was carried out on the cluster installed at Nagoya University. 
This work is in part supported by Nagoya‐Melbourne Joint Research Workshops Fund, provided by Nagoya University and the University of Melbourne.
G.N. is financially supported by JST SPRING, Grant Number JPMJSP2125.
K.I. is supported by JSPS KAKENHI Grant Number 24K00625 and the JST FOREST
Program JPMJFR20352935.
Finally, we thank Sreedhar Balu for fruitful discussions. 

\appendix

\section{Gas profiles in minihalos}
\label{sec:haloanalyticprof}
Here we introduce the analytical properties of halos to paste to the location of the halos in the simulation data. 
The gas density profiles $\rho_{\mathrm{g}}(r)$ is assumed that the gas is isothermal with virial temperature $T_{\vir}$ and in the hydrostatic equilibrium with the NFW DM profile \cite{2001PhR...349..125B, 1998ApJ...497..555M,2011MNRAS.410.2025X,1998ApJ...509..544S}, 
\begin{equation}
    \label{rhogas:eq1}
    \ln{\rho_{\mathrm{g}}(r)} = \ln{\rho_{g0}} - \frac{\mu m_\mathrm{p}}{2 k_{\mathrm{B}} T_{\mathrm{vir}}} \left[ v_{\mathrm{esc}}^2(0) - v_{\mathrm{esc}}^2(r) \right], 
\end{equation}
where the square of the escape velocity is 
\begin{equation}
    \label{vesc:eq}
    v^2_{\mathrm{esc}}(r) = 2 \int^{\infty}_{r}  \frac{GM(r')}{r'^2} dr' = 2V_\mathrm{c}^2 \frac{f(cx)+cx/(1+cx)}{xf(c)} 
\end{equation}
with $f(x) = \ln{(1+x)} - x/(1+x)$ and the concentration parameter $c \equiv R_\vir / r_\mr{s}$ by the scale radius $r_\mr{s}$ and the virial radius $R_\mr{vir}$, and circular velocity $V_c$: 
\begin{equation}
    \label{vcir:eq}
    V_{c} = \sqrt{\frac{GM}{R_{\mathrm{vir}}}} = 23.4 \left( \frac{M}{10^8 h^{-1} M_{\odot}} \right)^{1/3} \left[\frac{\Omega_m}{\Omega_m^z} \frac{\Delta_c}{18\pi^2}\right]^{1/6} \left( \frac{1+z}{10} \right)^{1/2} \;\; \mathrm{km/s},
\end{equation}
and the central gas density $\rho_{g0}$:  
\begin{equation}
    \label{rhogas:eq2}
    \rho_{g0}(z) = \frac{(\Delta_c/3) c^3 e^A}{\int_{0}^{c} (1+t)^{A/t} t^2 dt} \left( \frac{\Omega_b}{\Omega_m} \right) \Omega_m^z \rho_\mr{crit}(z),
\end{equation}
where $A = 2c/f(c)$, and the virial temperature $T_\mr{vir}$: 
\begin{equation}
    \label{tvir:eq}
    T_{\mathrm{vir}} = 1.98 \times 10^4 \left( \frac{\mu}{0.6} \right) \left( \frac{M}{10^8 h^{-1} M_{\odot}} \right)^{2/3}
    \left[\frac{\Omega_m}{\Omega_m^z} \frac{\Delta_c}{18\pi^2}\right]^{1/3} \left(\frac{1+z}{10} \right) \;\; \mathrm{K}.
\end{equation}

As for the concentration parameter, we use the fitting formula by \cite{2021MNRAS.506.4210I}. 
$R_{\mathrm{vir}}$ refers to the virial radius \cite{2001PhR...349..125B}, 
\begin{equation}
    \label{rvir:eq}
    R_{\mathrm{vir}} = 0.784 \left( \frac{M}{10^8 h^{-1} M_{\odot}} \right)^{1/3} \left[\frac{\Omega_m}{\Omega_m^z} \frac{\Delta_c}{18\pi^2}\right]^{-1/3}
    \left( \frac{1+z}{10} \right)^{-1} h^{-1} \;\; \mathrm{kpc}.
\end{equation}
Here, $\Delta_c = 18 \pi^2 + 82d -39d^2$ is the over density of the virialized halo collapsing at the redshift $z$ 
with $d = \Omega_\mr{m}^z - 1$ and $\Omega_m^z = \Omega_m (1+z)^3/(\Omega_m(1+z)^3 + \Omega_\Lambda)$.

\section{The overestimation of $\tau_\mr{halo}$ by the formula for $\tau_\mr{IGM}$}
\label{sec:tauhalooverestimate}
In this appendix we evaluate how the optical depth estimation is influenced if the equation (\ref{tauofIGM:eq}) is used 
even when the frequency resolution is finer than the minihalos. 

As explained in Subsection \ref{subsec:tauformula}, equation (\ref{tauofIGM:eq}) is basically allowed to be used where $n_\mr{HI}$, $T_\mr{S}$, $T_\mr{k}$ and $v_\mr{pec}$ can be regarded as constants in the integration range of equation (\ref{tauofmh:eq}). 
To estimate the optical depth from minihalos, we need enough resolution to resolve the halos. 
Because the pixel length is smaller than the diameter of the minihalos and because the FWHM of the minihalos is longer than the pixel length, the physical properties for the integration of equation (\ref{tauofmh:eq}) are not constant within the pixels located inside the FWHM of any halos. 

\begin{figure}[htbp]
\centering
\includegraphics[width=0.6\textwidth]{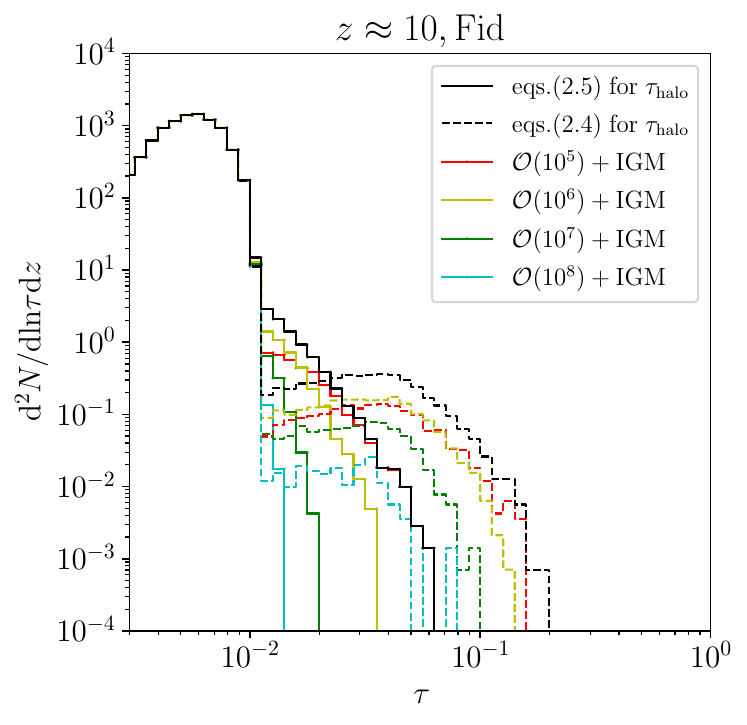}
\caption{
The abundance of the optical depth $\mrd N^2/ \mrd \ln \tau \mrd z$ at $z \approx 10$ in the fiducial model with the frequency resolution $\Delta \nu \approx 1 \mr{kHz}$.
The solid lines represent the results obtained using equation (\ref{tauofmh:eq}) to calculate the minihalo optical depth with the highest resolution whose scale is smaller than the any halos, and the dashed lines correspond those obtained using equation (\ref{tauofIGM:eq}).  
Different colors represent different halo mass ranges as indicated in the figure.
}
\label{fig:ETF}
\end{figure}

Figure \ref{fig:ETF} shows the optical depth abundance computed with both minihalos and the diffuse IGM at $z \approx 10$ with $\Delta \nu \approx 1 \mr{kHz}$. 
The solid lines are the optical depth distribution, in which the optical depth of minihalos is calculated using equation (\ref{tauofmh:eq}), the same method used for the other results in this paper. 
The dashed lines are the results in which the optical depth of minihalos are calculated using equation (\ref{tauofIGM:eq}). 
In this case, the optical depth contribution of each minihalo, and hence the total contribution, are overestimted.
The maximum optical depth is several times larger, and the number of larger optical depth is significantly enhanced. 
Therefore, when investigating the 21 cm forest including the minihalo influence, it is important to carefully evaluate the pixel scale compared to the halo scale, and to ensure the correct application of equations.

\bibliography{ref}
\bibliographystyle{JHEP}









\end{document}